\documentclass[a4paper,twocolumn,11pt]{quantumarticle}

\pdfoutput=1
\usepackage[utf8]{inputenc}

\usepackage[numbers,sort&compress]{natbib}

\usepackage[english]{babel}
\usepackage[T1]{fontenc}
\usepackage{csquotes}
\usepackage{xcolor}

\usepackage{amsmath}
\usepackage{amsfonts}
\usepackage{hyperref}
\usepackage{physics}
\usepackage{subcaption}
\usepackage{tikz}
\usepackage{lipsum}

\begin{document}

\title{Towards molecular docking with neutral atoms}

\author{Mathieu Garrigues}
\email{mathieu.garrigues@protonmail.com}
\affiliation{Independent researcher}
\orcid{0009-0008-8630-5177}

\author{Victor Onofre}
\email{vonofre68@gmail.com}
\affiliation{Quantum Open Source Foundation}
\orcid{0000-0002-6175-9171}

\author{Noé Bosc-Haddad}
\email{noe.bosc.haddad@gmail.com}
\affiliation{CentraleSupélec,Gif-Sur-Yvette,France}

\thanks{\vspace{1.5em}}

\thanks{\\ Mathieu Garrigues: These authors contributed equally to this work}

\thanks{\\ Victor Onofre: These authors contributed equally to this work}

%\thanks{You can use the \texttt{\textbackslash{}email}, \texttt{\textbackslash{}homepage}, and \texttt{\textbackslash{}thanks} commands to add additional information for the preceding \texttt{\textbackslash{}author}. If applicable, this can also be used to indicate that a work has previously been published in conference proceedings.}

\maketitle

\begin{abstract}
%New computational strategies are emerging that can speed up the drug discovery process, such as Virtual screening (VS) which highlights molecules that fit the target. VS uses structure-based methods such as molecular docking to predict the activity of candidate molecules at the binding site of the proteins. This helps in selecting those molecules that exhibit desirable behavior while rejecting the rest. However, for large chemical libraries, it is necessary to search and score configurations using the least computational resources but still with high precision. 
New computational strategies, such as molecular docking, are emerging to speed up the drug discovery process. This method predicts the activity of molecules at the binding site of proteins, helping to select the ones that exhibit desirable behavior and rejecting the rest. However, for large chemical libraries, it is essential to search and score configurations using fewer computational resources while maintaining high precision. 

In this work, we map the molecular docking problem to a graph problem, a maximum-weight independent set problem on a unit-disk graph in a physical neutral atom quantum processor. Here, each vertex represents an atom trapped by optical tweezers. The Variational Quantum Adiabatic Algorithm (VQAA) approach is used to solve the generic graph problem with two optimization methods, Scipy and Hyperopt. Additionally, a machine learning method is explored using the adiabatic algorithm. Results for multiple graphs are presented, and a small instance of the molecular docking problem is solved, demonstrating the potential for near-term quantum applications.

\end{abstract}

%%-----------------------------------------------------%%

\section{Introduction}

% At the moment, a lot of innovation is being done to develop quantum algorithms that could bring an advantage over classical computers using only Noisy Intermediate Scale Quantum (NISQ) devices. These advantages span from more accurate results to faster algorithms and lower energy consumption. Among the different quantum hardwares and methodologies, neutral atoms are particularly well suited to solve combinatorial graph problems. In the neutral atoms devices, adiabatic algorithms or similar methods can be applied in a natural way, the Ising Hamiltonian describing the dynamics of the qubits is closely related to the cost function to be minimized, as has been shown in the case of the Maximum Independent Set (MIS) problem \cite{wurtz2022industry, coelho2022efficient}.

In the quest for faster and more efficient drug discovery, new computational strategies are emerging, among them, molecular docking. This method helps predict how molecules interact with the binding sites of proteins, which in turn helps in selecting promising candidates while discarding others. However, with vast chemical libraries, finding and evaluating configurations swiftly and accurately remains a challenge.

Quantum computing has become a focal point for its potential to revolutionize optimization tasks, which are prevalent across industries. In this paper, we delve into the fusion of quantum computing and optimization, specifically exploring the use of neutral atom devices for tackling the combinatorial optimization problem that is molecular docking.

Neutral atom devices present a promising avenue for quantum computing, boasting extended coherence times and precise control over individual atoms. By trapping and manipulating atoms with laser fields, these devices enable quantum gate operations and system simulations. Leveraging these capabilities, we investigate their efficacy in solving intricate optimization challenges.

Our research focuses on employing neutral atom devices to tackle the Maximum Independent Set (MIS) problem \cite{wurtz2022industry, coelho2022efficient}, fundamental in various real-world applications. The approach involves encoding graph representations into atomic quantum states, implementing quantum adiabatic evolution, and using machine learning to expedite parameter optimization.

This paper unfolds as follows: we lay the theoretical groundwork by discussing the binding interaction graph model and the Maximum Clique Problem (MCP). Then, we introduce neutral atom devices, explaining their operational principles and suitability for quantum computation. Next, we detail our methodology, covering optimization problem mapping, register formation with neutral atoms, and the integration of quantum links for non-local interactions. We also delve into the Quantum Adiabatic Algorithm (QAA) and its variant, the Variational Quantum Adiabatic Algorithm (VQAA), elucidating their roles in solving the MIS problem. Additionally, we introduce machine learning techniques to replace and accelerate parameter optimization in the VQAA approach. Finally, we analyze the results from our simulations and optimizations, shedding light on the effectiveness and limitations of our proposed methodologies.

%%-----------------------------------------------------%%

\section{Contribution}

In this work, we implement the Neutral Atoms VQAA algorithm to identify the Maximum Independent Set of a graph in the context of molecular docking. Furthermore, we propose an enhancement to this algorithm by using two different classical optimizers and leveraging a graph machine learning approach to significantly accelerate its computational efficiency.

%In the spirit of promoting reproducibility and transparency,
The codebase utilized for obtaining the results presented in this study is openly available on our \href{https://github.com/M-Garrigues/Quantum-Docking}{GitHub repository}. Researchers interested in exploring the implementation details, reproducing the experiments, or building upon our work can use it freely.

%%-----------------------------------------------------%%

\section{Molecular docking}

Virtual Screening (VS) is a computational technique that employs sophisticated algorithms and molecular modeling to identify and prioritize potential drug candidates from vast libraries of chemical compounds \cite{halperin2002principles, agu2023molecular}.

\begin{figure}[ht!]    
\centering
\includegraphics[width=0.9\linewidth]{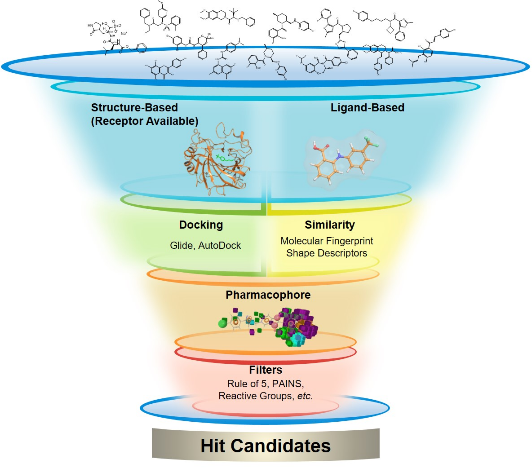}  
    \caption{Virtual Screening process pipeline. A ligand database goes through multiple filters in order to identify Hit Candidates. Image from \cite{VSimage}.
}
    \label{fig:virtual_screening}
\end{figure}

This approach serves as a pivotal tool for researchers and pharmaceutical companies in streamlining the drug development process by significantly narrowing down the list of compounds that need to be synthesized and tested experimentally \cite{pinzi2019molecular}. Molecular Docking is a technique used in Virtual Screening (VS) that tries to predict the interactions between a drug molecule (ligand) and a target protein molecule (receptor). Docking algorithms can be classified in 2 main groups, rigid-body and flexible docking \cite{morris2008moleculardocking}. Rigid-body docking algorithms like ZDOCK \cite{chen2003zdock}, relies mainly on the geometrical features of the ligand and receptor. Flexible docking on the other hand can consider different arrangements of the ligand and/or receptor which comes with a higher computational cost. Some flexible docking methods are for example Fast Shape Matching \cite{kuntz1982fastshapematching}, Monte Carlo simulation \cite{liu1999montecarlodock}, Distance Geometry \cite{more1999distancegeometrydocking}.
The output of this techniques is the predicted three-dimensional orientation of the ligand with respect to the receptor binding site, along with a corresponding score for each orientation.
For an accurate determination of the most probable ligand orientation and its ranking among other compounds, precise scoring functions and efficient search algorithms are necessary. The scoring function is a set of physical or empirical parameters that are used to score the binding orientation and interactions, based on experimentally determined data on active and inactive ligands. It will be detailed in section \ref{sec:scoring_function}.

%%-----------------------------------------------------%%

\section{Molecular docking as a graph problem}
\label{sec:molecular_docking_as_graph}

\subsection{Molecule to graph} \label{subsec:4.1} 

\begin{figure}[ht!]
\centering
    \includegraphics[width=0.4\linewidth]{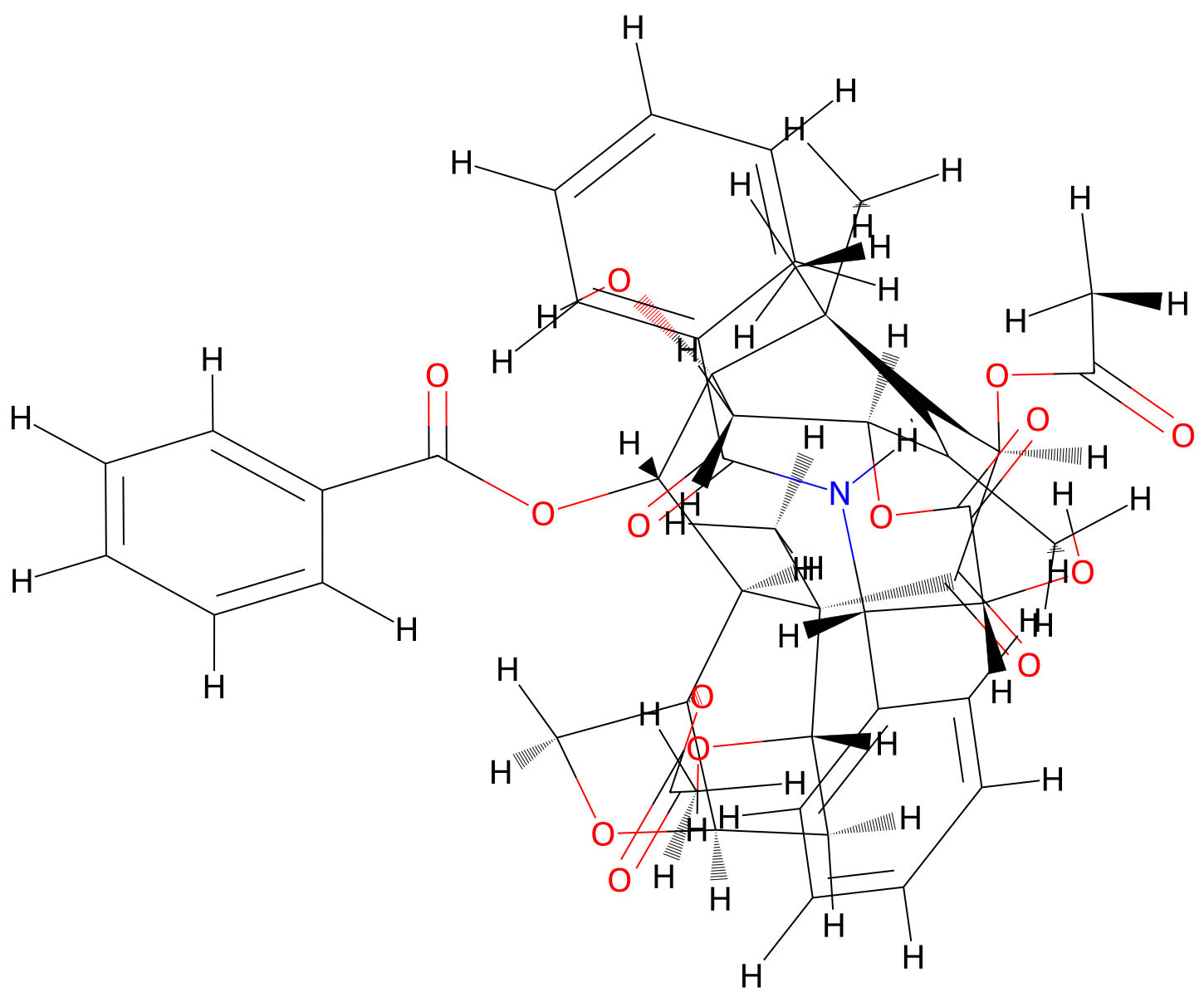}  
    \includegraphics[width=0.4\linewidth]{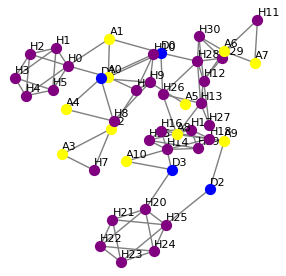}  
    \caption{(Left) Planar structure of the Taxol ligand molecule. (Right) Full graph of ligand molecule pharmacophore points.}
    
    \label{fig:molecule_taxol}
\end{figure}

\begin{figure}[ht!]
\centering
    \includegraphics[width=0.5\linewidth]{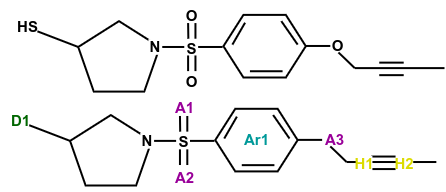}  
    \includegraphics[width=0.5\linewidth]{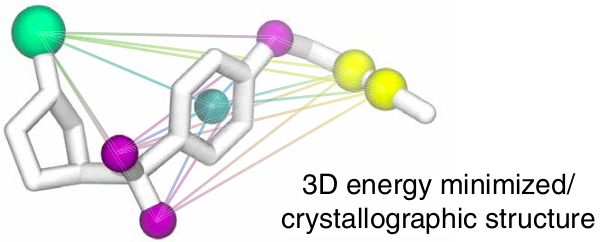}  
    \caption{(Top) Planar structure of example molecule with their pharmacophore points in colors letters and (Bottom) 3D structure of the same molecule. Image from \cite{banchi2020molecular}.}
    \label{fig:3d_molecule}
\end{figure}

We follow the process presented in \cite{banchi2020molecular} to map the molecule to a graph. The first step is to look at the planar structure of the molecule to identify the pharmacophore points. An example is shown for the Taxol molecule in figure \ref{fig:molecule_taxol} and and more simple structure in figure \ref{fig:3d_molecule}.

\begin{figure}[ht!]
\centering
    \includegraphics[width=0.45\linewidth]{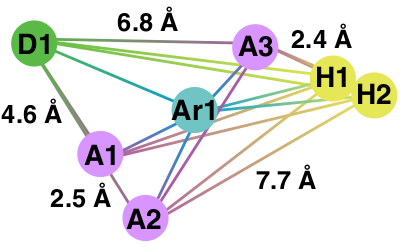}  
    \caption{Graph representation of molecule shown in figure \ref{fig:3d_molecule}. Image from \cite{banchi2020molecular}.}
    \label{fig:molecule_graph}
\end{figure}

With the 3D structure (figure \ref{fig:3d_molecule}) of the molecule, we can determine the pairwise distance between the pharmacophore points. This information represents a molecule as a graph with weighted edges as shown in figure \ref{fig:molecule_graph}.

\begin{figure}[ht!]
\centering
    \includegraphics[width=0.8\linewidth]{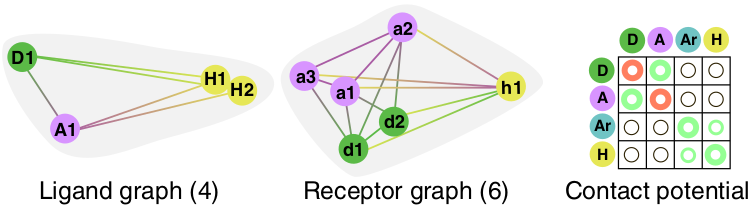}  
    \caption{Ligand and receptor graphs with the corresponding contact potential. Image from \cite{banchi2020molecular}}
    \label{fig: graph to binding graph A}
\end{figure}

\begin{figure}[ht!]
\centering
    \includegraphics[width=0.8\linewidth]{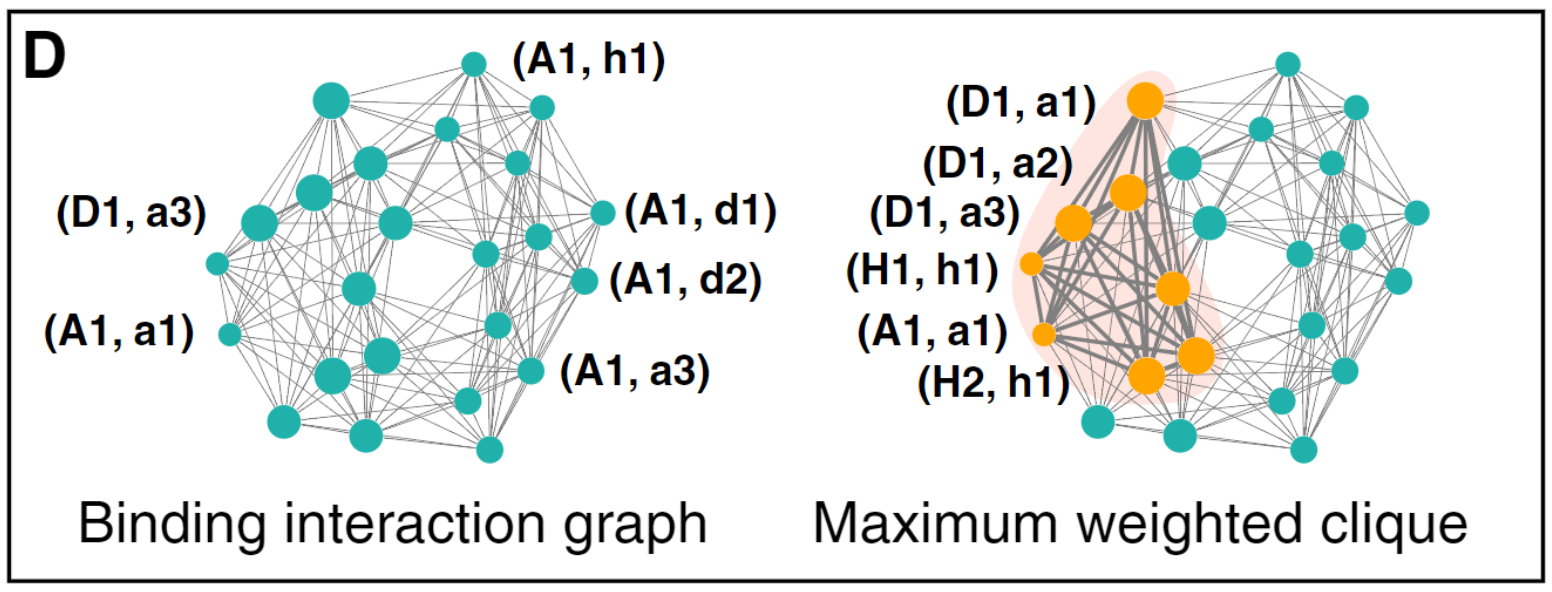}  
    \caption{Example of binding interaction graph and the maximum clique found. Image from \cite{banchi2020molecular} }
    \label{fig:full_map_binding_interaction}
\end{figure}

%%-----------------------------------------------------%%
\subsection{Binding interaction graph} \label{subsec:4.2}
%%-----------------------------------------------------%%
In order to model possible binding poses, the protein and ligand are represented as a graph. This is done by following the steps presented in the previous section then by constructing the binding interaction graph which represents the sets of contacts. 

The different types of pharmacophore features have a specific attraction force, if any, between one another. If one feature point from the ligand and one from the receptor are attracted to each other, they form a potential contact point (figure \ref{fig: graph to binding graph A}). These points are the nodes of the binding interaction graph. The nodes are assigned weights based on the distance between the features and their mutual interaction strength. 
Two contacts are mutually compatible if their mutual realization do not violate the geometrical shapes of the ligand and the binding site. In this case, an edge is added between these compatible points in the binding interaction graph. Therefore, a pairwise compatible set of contacts that arise from a true binding pose forms a complete subgraph of the binding interaction graph \cite{banchi2020molecular}. The obtained graph is shown in figure \ref{fig:full_map_binding_interaction}.

A complete subgraph, also known as a clique, is a subgraph in a graph G where all possible pairs of vertices are connected by an edge.

%%-----------------------------------------------------%%
\subsection{Maximum Clique Problem} \label{subsec:4.3}
%%-----------------------------------------------------%%

A $G=(V, E)$ is an arbitrary undirected and weighted graph, where $V=\{1,2, \ldots, n\}$ is the vertex set of $G$ (we use the terms vertex and node synonymously throughout), and $E \subseteq V \times V$ is the edge set of $G$. 
The complement graph of $G=(V, E)$ is the graph $\bar{G}=(V, \bar{E})$, where $V=\{1,2, \ldots, n\}, \bar{E}=\{(i, j) \mid i, j \in V, i \neq j,(i, j) \notin E\}$. 

The concept of the Max Clique problem is intimately linked to the determination of binding conformations and the energetic considerations governing ligand-receptor interactions.
% When constructing pharmacophore graphs for both the ligand and the receptor, these graphs are essentially representations of the spatial arrangements of essential features within the molecules. 

It is a well-known graph theory problem that seeks to find the largest complete subgraph within a given graph, where every node is connected to every other node. Finding the max clique in the binding interaction graph implies identifying the largest arrangement of pharmacophore features that can interact simultaneously with the receptor.

However, when identifying binding conformations, it is equally important to assess the number of connections, but also their energy. To identify correctly probable poses, the total interactions energy between the ligand and the receptor must be taken into account. The problem is therefore transformed in a Weighted Max Clique Problem (WMCP). In essence, the WMCP identifies the binding conformation that not only adheres to geometric criteria but also maximizes the weighted sum of interactions, focusing on the most pivotal interactions within the ligand-receptor complex.

For each vertex $i \in V$, a positive weight $w_i$ is associated with $i$, collected in the weight vector $w \in \mathbb{R}^n$.
The WMCP  can then be defined as follows:
\begin{equation}
\begin{gathered} \max \sum_{i=1}^n w_i x_i, \\ \text { s.t. } x_i+x_j \leq 1, \forall(i, j) \in \bar{E}, \\ x_i \in\{0,1\}, i=1, \ldots, n . \end{gathered}
\end{equation}

Because of the nature of neutral atom devices, they are not suited to solve the Maximum Clique Problem. However they are naturally capable of solving the Maximum Weighted Independent Set problem which can be mapped directly from Maximum Clique Problem by working on the complementary graph as seen in figure \ref{fig:MIS to Max clique}.

%%-----------------------------------------------------%%
\subsection{Scoring Function} \label{sec:scoring_function}
%%-----------------------------------------------------%%

A classical WMCP heuristic is generally ran multiple times, producing a list of probable conformations, which are then evaluated with a scoring function to identify the actual best ones. 

Typically, scoring functions consider a combination of geometric and energetic factors, but more precise (and compute-heavy) than the ones used when assigning weights during the interaction graph generation \cite{halperin2002principles}. Geometric terms evaluate how well the ligand fits within the receptor's binding site, considering factors such as steric clashes and complementarity of shape. Energetic terms account for the interaction energies between the ligand and the receptor, encompassing contributions from van der Waals forces, electrostatic interactions, and hydrogen bonding \cite{warren2006critical}. The scoring function quantifies these factors and assigns a score to each binding conformation, enabling the identification of the most likely and energetically favorable binding modes.

%%-----------------------------------------------------%%

\section{Neutral atoms devices}

\begin{figure*}
\centering
\includegraphics[width=0.95\linewidth]{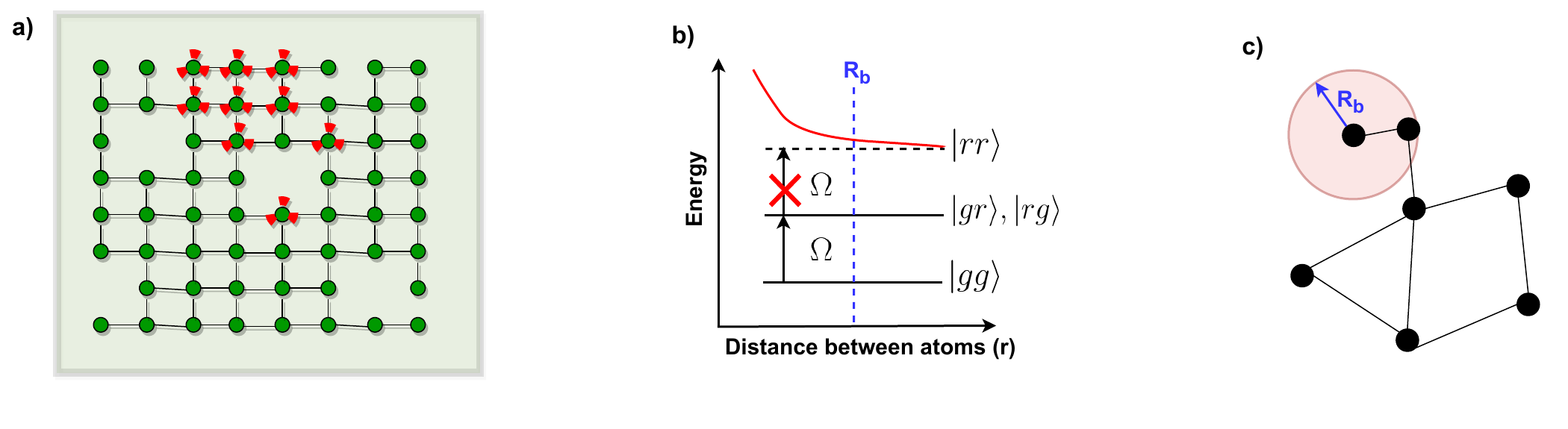}  
    \caption{a) Arrays of optical tweezers are used to prepare a register made of neutral atoms . b) Rydberg blockade Effect, an atom cannot be excited to the Rydberg level if a nearby atom is already in such state. c) Rydberg atoms correspond to the nodes of a UD-graph.}
    \label{fig:pasqal_devices}
\end{figure*}

Neutral atom devices are made up of two main components: the register and the channels \cite{foot2004atomic}. The register is a group of trapped atoms arranged in a defined but changeable configuration. Each atom holds a specific quantum state that is encoded in a particular electronic level (as shown in figure \ref{fig:pasqal_devices}.a). On the other hand, the channels are responsible for manipulating the state of the atoms by addressing specific electronic transitions. These channels consist mostly of lasers. Each channel is tuned such that each of its pulses coherently drives a specific electronic transition between two energy levels of an atom. 

%For instance, when addressing the ground-rydberg transition, we have that our targeted levels are the ground state, $\ket{g}$ , and the Rydberg state, $\ket{r}$ . These two states can be thought of as the two levels of a quantum spin.

In this system, a pulse acting on a atom $i$, with Rabi fequency $\Omega(t)$, detuning $\delta(t)$ and a fixed phase $\phi$, will have the Hamiltonian terms:

\begin{equation}
\frac{\hbar \Omega(t)}{2} (\cos{(\phi)}) \sigma^x_{i} - \sin{(\phi)}  \sigma^y_{i}) - \frac{\hbar}{2} \delta(t) \sigma^z_{i} 
\label{eq:hamiltonian_1}
\end{equation}

where $\sigma^\alpha$ for $\alpha = x,y,z$ are the Pauli matrices.

%Alternatively, once can rewrite this term as: 

%\begin{equation}
%\frac{\hbar}{2}  \mathbf{\Omega(t)} \cdot %\mathbf{\sigma}_{i}, 
%\label{eq:hamiltonian_2}
%\end{equation}

%where $\mathbf{\Omega(t)} = (\Omega(t) \cos{\phi}, -\Omega(t) \sin{\phi}, - \delta(t)  ) ^T $ and $\mathbf{\sigma}$ is the vector of Pauli matrices.

Atoms in neutral atom devices can be driven to Rydberg states to enable them to interact over large distances. The Van der Waals force describes the interaction between two atoms at the same Rydberg level and at a distance $R$, which scales as $R^{-6}$. This interaction can be leveraged to create fast and reliable quantum gates using the Rydberg blockade Effect.

%This effect consists on the shift in energy between the doubly excited Rydberg state of nearby atoms and their ground state, making it non-resonant with an applied laser field coupling the ground and Rydberg levels. 

The Rydberg blockade is a phenomenon where an atom cannot be excited to the Rydberg level if there is already another atom nearby in such state(as shown in figure \ref{fig:pasqal_devices}.b). In order to represent this interaction, we consider it as a penalty term in the Hamiltonian that describes the state where both the atoms are excited: $U_{ij} n_i n_j$, where $n = (1 + \sigma^z) / 2 $ is the projector on the Rydberg state, $U_{ij}  \propto R^{-6}_{ij}$ and $R_{ij}$ is the distance between the atoms $i$ and $j$. 

%\begin{equation}
%U_{ij} n_i n_j,
%\label{eq:hamiltonian_condition}
%\end{equation}

The proportionality constant is set by the chosen Rydberg level. An entire array of interacting atoms acted on by the same pulse can be represented as an Ising-like Hamiltonian:

\begin{equation}
H = \frac{\hbar}{2} \sum_i \Omega(t) \sigma^x_{i} - \frac{\hbar}{2}  \sum_i \delta(t) \sigma^z_{i}  + \sum_{i<j} U_{ij} n_i n_j     , 
\label{eq:hamiltonian_3}
\end{equation}

In the analog quantum simulation approach, the laser field acts on the entire array of atoms. This creates a global Hamiltonian of the form in equation \ref{eq:hamiltonian_3}.

%\begin{equation}
%H = \sum_i \Big(   \frac{\hbar \Omega(t) }{2}   \sigma^x_{i} - \frac{\hbar \delta(t)}{2}   \sigma^z_{i}  + \sum_{i<j} U_{ij} n_i n_j   \Big) 
%\label{eq:hamiltonian_3}
%\end{equation}

Through the continuous manipulation of $\Omega(t)$ and $\delta(t)$, one has a very high degree of control over the system's dynamics and properties. In this way, the analog approach enables the quantum simulation of many body quantum systems, but also provides novel ways of solving combinatorial problems that can be mapped onto the Hamiltonian above.

With \textit{Pulser} we can simulate the behavior of neutral atom devices. \textit{Pulser} is an open-source Python software package from Pasqal. It provides easy-to-use libraries for designing and simulating pulse sequences that act on programmable arrays of neutral atoms \cite{pulser}.

\begin{figure}[h] %hbt!
\centering
\includegraphics[width=0.95\linewidth]{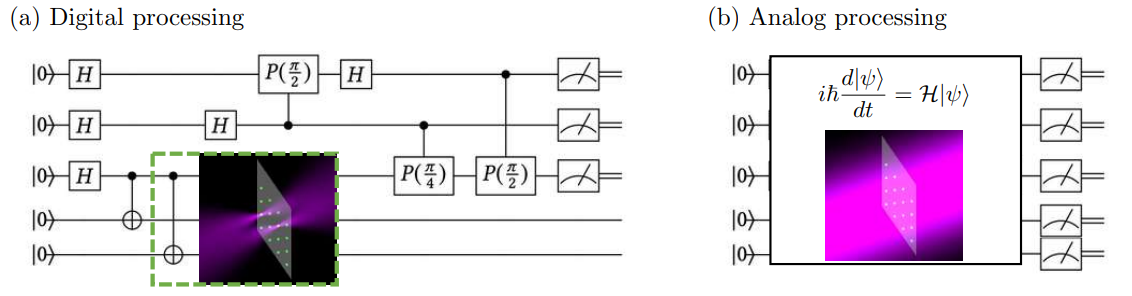}  
\caption{(a) In digital processing, a succession of discrete consecutive steps, gates, is applied to the qubits to implement the quantum evolution. (b) In analog processing the qubits evolve as the result of a time-dependent control of the Hamiltonian acting upon the qubits. Image from \cite{henriet2020quantum}}
\label{fig:digital_analog}
\end{figure}

%%-----------------------------------------------------%%

\section{Maximum-weight independent sets}

An independent set of a graph $G$ is the subset of vertices $S \subseteq  V$, such that none of the vertices in $S$ are connected by an edge in $G$. The largest such independent set is called a maximum independent set. The problem of finding a MIS is called the maximum independent set problem. 

The MIS problem can be generalized to the maximum-weight independent set problem, where each vertex is assigned a weight $w_{i}$, and accordingly, a weight $W_{S}$ is assigned to each subset of vertices $S \subseteq V$ via $W_{S} = \sum_{i \in S} w_{i}$ The MWIS problem is to find an independent set with the largest weight. It can be formulated as an energy minimization problem. For this, one can associate a binary variable $z_{i} \in {0,1}$ with each vertex $i \in V$. This allows us to identify a subset of vertices $S$ by a bitstring $z = (z_{1},..., z_{n})$ via $S = {i \in V | z_{n} = 1}$. Using this representation, we can consider the cost function

\begin{equation}
 C_{MWIS}(z_{1},...,z_{N}) = - \sum_{i}^{N} w_{i} z_{i} +  \sum_{(i,j)\in E} u_{ij} z_{i} z_{j}
\label{eq:cost_function}
\end{equation}

\begin{equation}
H = \frac{\hbar}{2} \sum_i \Omega(t) \sigma^x_{i} - \frac{\hbar}{2}  \sum_i \delta(t) \sigma^z_{i}  + \sum_{i<j} U_{ij} n_i n_j
\label{eq:neutral_atoms_and_MIS}
\end{equation}

The last two terms of the neutral atoms Hamiltonian (eq. \ref{eq:neutral_atoms_and_MIS}) have the same form as the cost function of the MWIS problem (eq. \ref{eq:cost_function}). At $\Omega = 0$, both equations coincide exactly, where each atom is placed at the respective location of the corresponding vertex, the blockade radius is identified with the unit disk radius and the weight of each vertex is identified with the local detuning $\delta_{i}$. As in the unweighted case, the ground state can be degenerate, corresponding to multiple independent sets achieving the same maximum weight. 

The MWIS set problem has applications spanning many disciplines, including signal transmission, information retrieval, and computer vision \cite{lamm2019exactly}.

%%-----------------------------------------------------%%

\section{Graph encoding in neutral atoms }

One of the features of neutral atom devices is that the ground state of the Hamiltonian can encode exactly the solution to MIS  on Unit-Disk (UD) graphs \cite{nguyen2023quantum}. A UD graph is a graph $G = (V,E)$ with vertices $V$ and edges $E$ that can be embedded in the 2D Euclidean plane such that has an edge between any two vertices whose Euclidean distance is less than 1. 

We are interested in unit-disk graphs since they are in one-to-one correspondence with atom arrangements in 2D. Specifically, each atom represents a vertex, and the low-energy states of the Hamiltonian do not contain states with pairs of atoms that are both in the Rydberg state if they are within some characteristic distance, defined as the blockade radius $r_{B}$ (as shown in figure \ref{fig:pasqal_devices}.c). This effect naturally imposes the independent set constricting on the ground state of the Hamiltonian at $\Omega = 0$, which allows one to encode the MWIS of corresponding UDG.

\begin{figure*}
\centering
\includegraphics[width=0.95\linewidth]{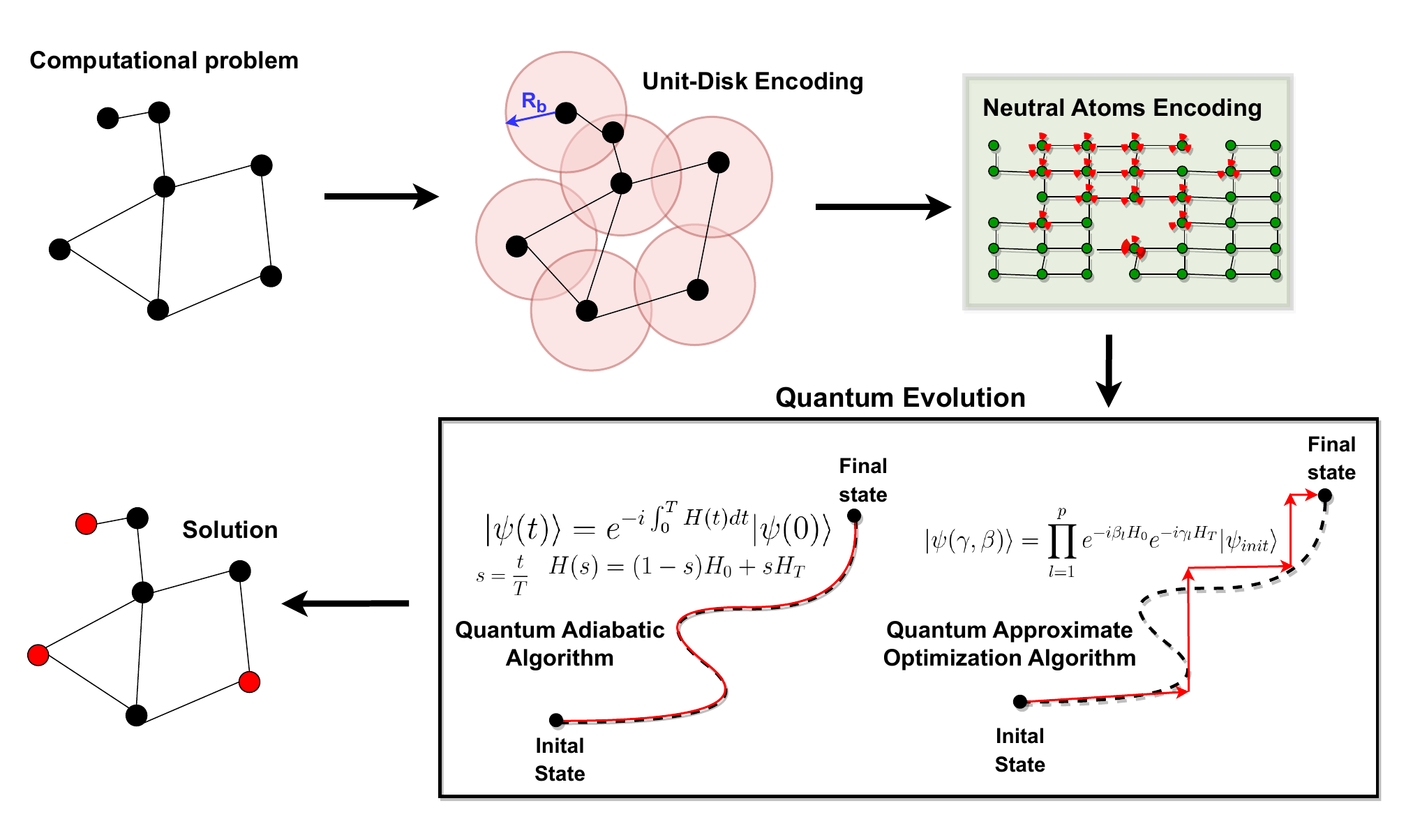}  
    \caption{Given the computational problem it is necessary to map it to a unit-disk graph for the encoding into the neutral atoms device. The most natural quantum evolution of the system is the QAA or QAOA.}
    \label{fig:workflow_process}
\end{figure*}

%%-----------------------------------------------------%%

\section{Methodology}

\label{sec:proposed_algorithm_section}
\subsection{Mapping to MWIS}

\begin{figure}[hbt!]
\centering
    \includegraphics[width=0.7\linewidth]{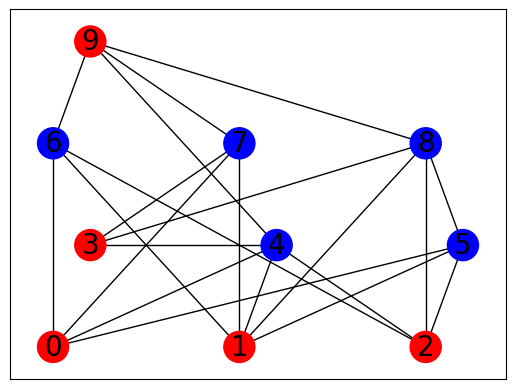}  
    \includegraphics[width=0.7\linewidth]{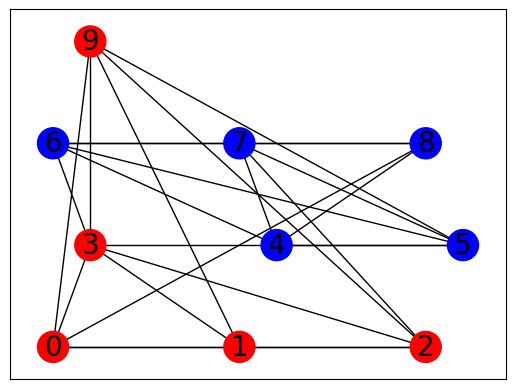}  
    \caption{(upper figure) The Maximum independent Set solution (lower figure) The Maximum Clique solution in the complementary graph.}
    \label{fig:MIS to Max clique}
\end{figure}

To map the problem to MWIS we can take the binding interaction graph constructed through the procedure explained in \ref{subsec:4.1} and \ref{subsec:4.2}.  The mapping from MWCP to MWIS, can be done easily by considering the complementary graph $\bar{G}=(V, \bar{E})$. Solving the MWIS problem on a graph is equivalent to solving the MWCP on its complementary graph as shown in  figure \ref{fig:MIS to Max clique} . Hence performing MWIS on the complementary graph of the binding interaction graph gives directly the maximum clique in the complementary graph, as the solution to each problem is composed of the same nodes.

\begin{figure*}
\centering
\includegraphics[width=0.95\linewidth]{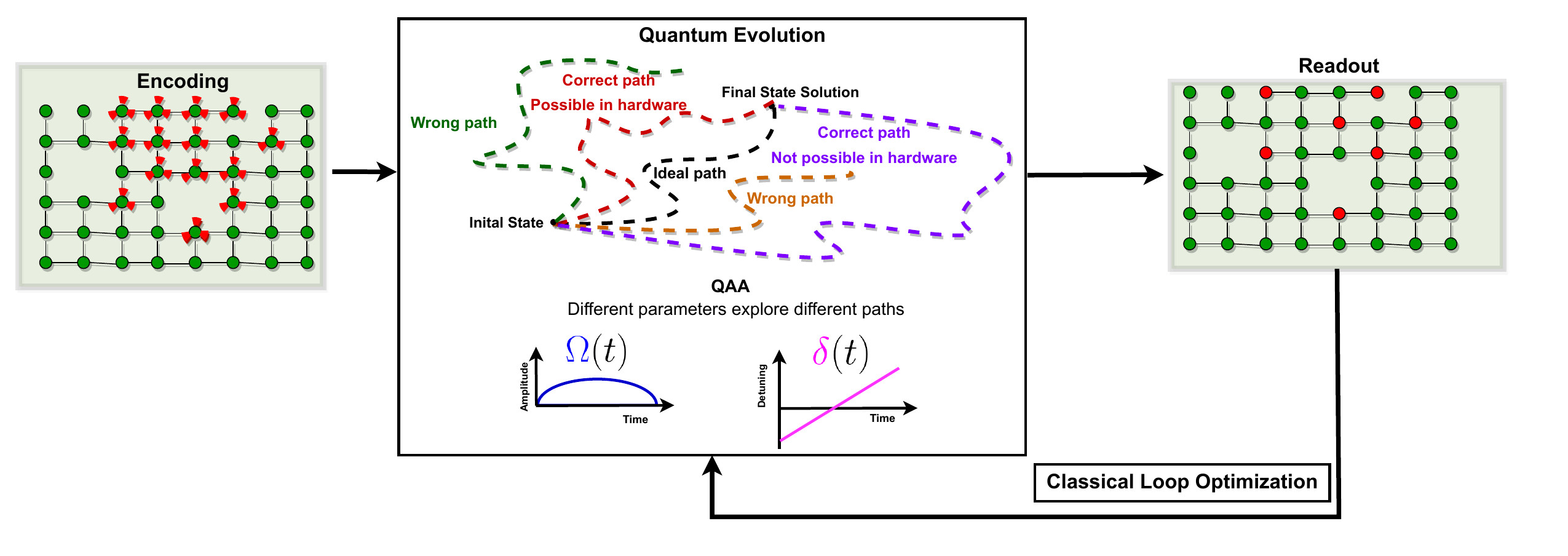}  
    \caption{VQAA worflow for solving the MIS problem with Neutral atoms. The VQAA allows for a significant acceleration compared to the QAA \cite{schiffer2022adiabatic}, yet requires fewer parameters and measurements than the QAOA. We aim at finding an
optimized path for the adiabatic evolution.}
    \label{fig:vqaa_process}
\end{figure*}

\subsection{Register formation}

\begin{figure}[hbt!]
\centering
      \includegraphics[width=0.5\linewidth]{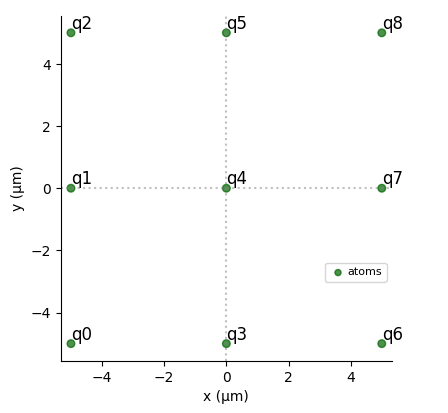}  
  \caption{Example of positions definitions of the atoms and their names using \textit{Pulser}. Each point represents an atom in the Neutral atom device.}
  \label{fig:example_from_register}
\end{figure}

\begin{figure}[hbt!]
\centering
      \includegraphics[width=0.95\linewidth]{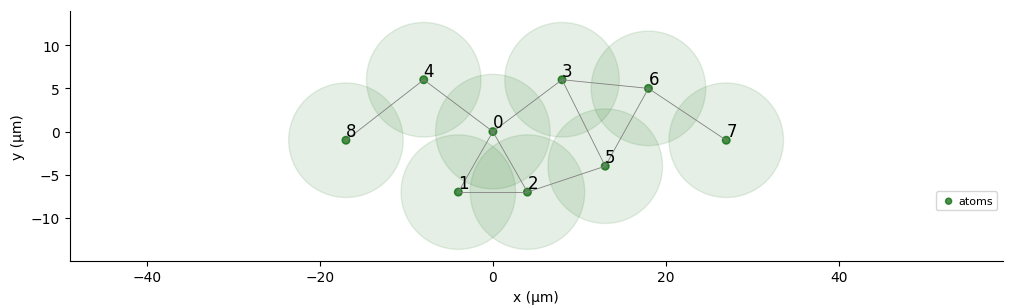}  
  \caption{Example of graph encoding using the unit-disk approach with \textit{Pulser}.}
  \label{fig:example_from_graph}
\end{figure}

To prepare a register made of neutral atoms, one can use arrays of optical tweezers \cite{henriet2020quantum}, we can use \textit{Pulser} to simulate this behavior. We can define an array of atoms to use as our register as shown in figure \ref{fig:example_from_register}. As we can observe in figure \ref{fig:example_from_graph}, each atom can represent a vertice in a graph using the Unit-disk approach. 

If we want to work on the evolution of a system with neutral atoms, the sequence (as named in \textit{Pulser}) is the central object and it essentially consists in a series of pulses (and other instructions) that are sequentially allocated to channels.

Each pulse describes, over a finite duration, the modulation of a channel’s output amplitude, detuning and phase. While the phase is constant throughout a pulse, the amplitude and detuning are described by waveforms, which define these quantities values throughout the pulse. In figure \ref{fig:example_adiabatic} we have an example of a pulse representing an adiabatic evolution of a graph to solve the MIS problem. 

\subsection{Quantum links} \label{sec:links}
In the common scenario, as often encountered in our use case, where the obtained graph exhibits non-local characteristics, featuring overlapping edges, the mapping to an UDG becomes complex. UDGs are fundamentally based on nearest neighbors, which virtually precludes the existence of non-local interactions.

\begin{figure}[h]
\centering
  \includegraphics[width=0.48\linewidth]{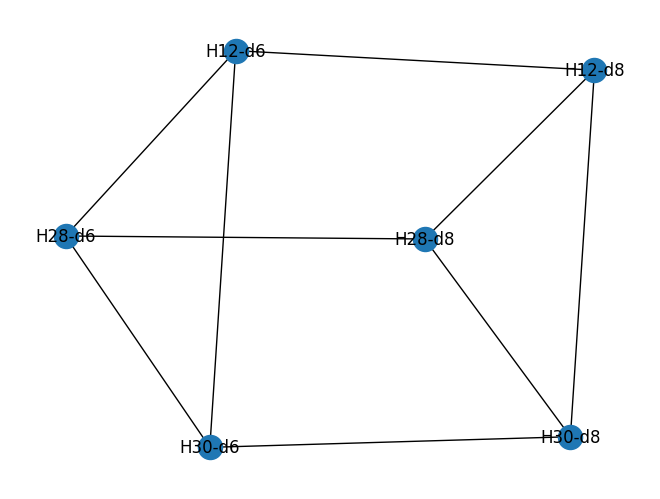}
  \includegraphics[width=0.48\linewidth]{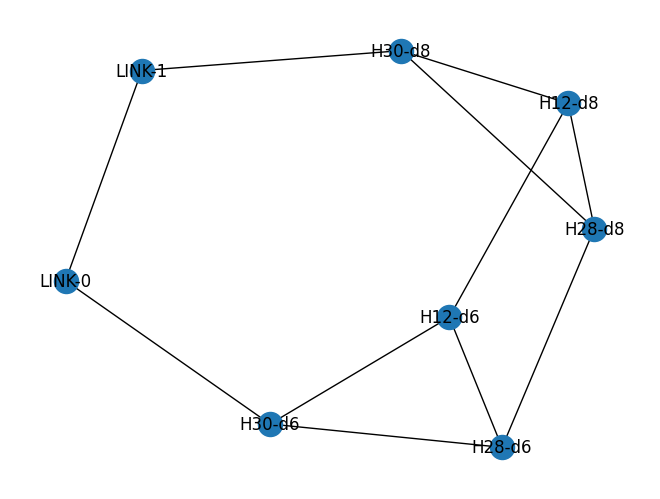}
  \label{fig:sub2}
\caption{Simple interaction graph with overlapping edges. On the right we add a quantum link between H30-d8 and H30-d6}
\label{fig:overlapping_edges}
\end{figure}

A straightforward solution to this challenge involves the use of quantum links\cite{quantum_wires_byun}. The concept is to introduce ancillary nodes between two distant yet connected nodes to physically link them through a chain of nodes, thereby entangling them. In the case of the MIS, the node chains should be created in pairs of ancillary nodes to preserve the system's state and ultimately yield the correct result. It is essential to disregard these quantum links during the result retrieval process. See examples with figure \ref{fig:overlapping_edges}.

\subsection{Quantum Adiabatic Algorithm (QAA)}

\begin{figure}[h]
\centering
      \includegraphics[width=0.8\linewidth]{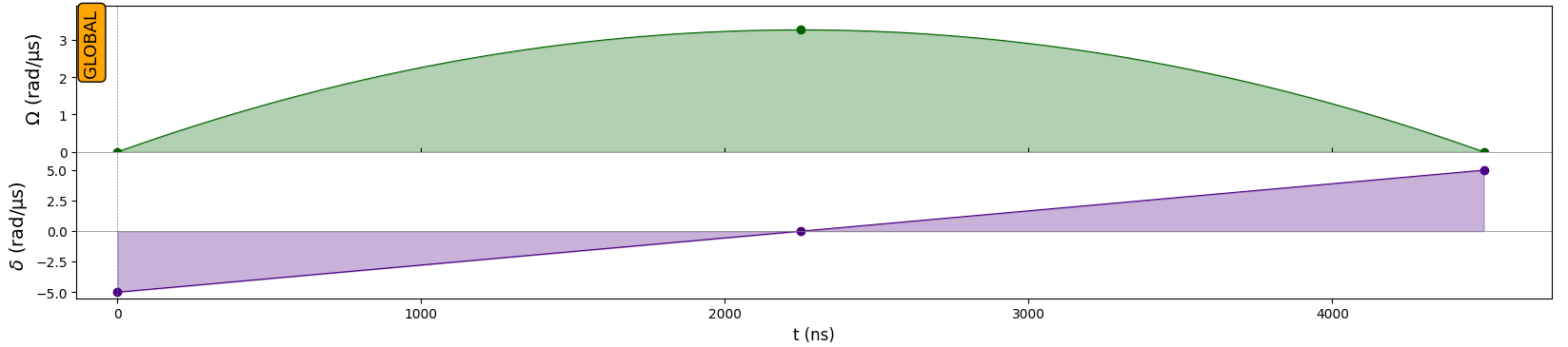}  
  \caption{Example of an adiabatic pulse for solving the MIS problem using \textit{Pulser}.}
  \label{fig:example_adiabatic}
\end{figure}

A method to solve the MIS problem is the Quantum Adiabatic Algorithm (QAA) \cite{albash2018adiabatic}. We consider a Hamiltonian of the following form:

\begin{equation}
H(t) = u(t) H_{M} + (1 - u(t) ) H_{C},
\label{eq:hamitonian_adiabatic}
\end{equation}

where $H_{C}$ is the problem (cost) Hamiltonian, it encodes the optimization task that we are trying to solve, $H_{M}$ is the mixer Hamiltonian, and encodes quantum mixing (a uniform transverse field on qubits), and $u(t)$ and the control function.
The idea behind the adiabatic algorithm  is to slowly evolve the system from an easy-to-prepare ground state to the ground state of $H$. If done slowly enough, the system of atoms stays in the instantaneous ground-state. An example of an Pulse applying the adiabatic method with \textit{Pulser} is showed in figure \ref{fig:example_adiabatic}.

\begin{figure}[h]
\centering
      \includegraphics[width=0.6\linewidth]{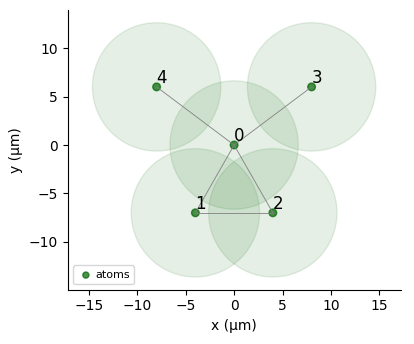}  
      \includegraphics[width=1\linewidth]{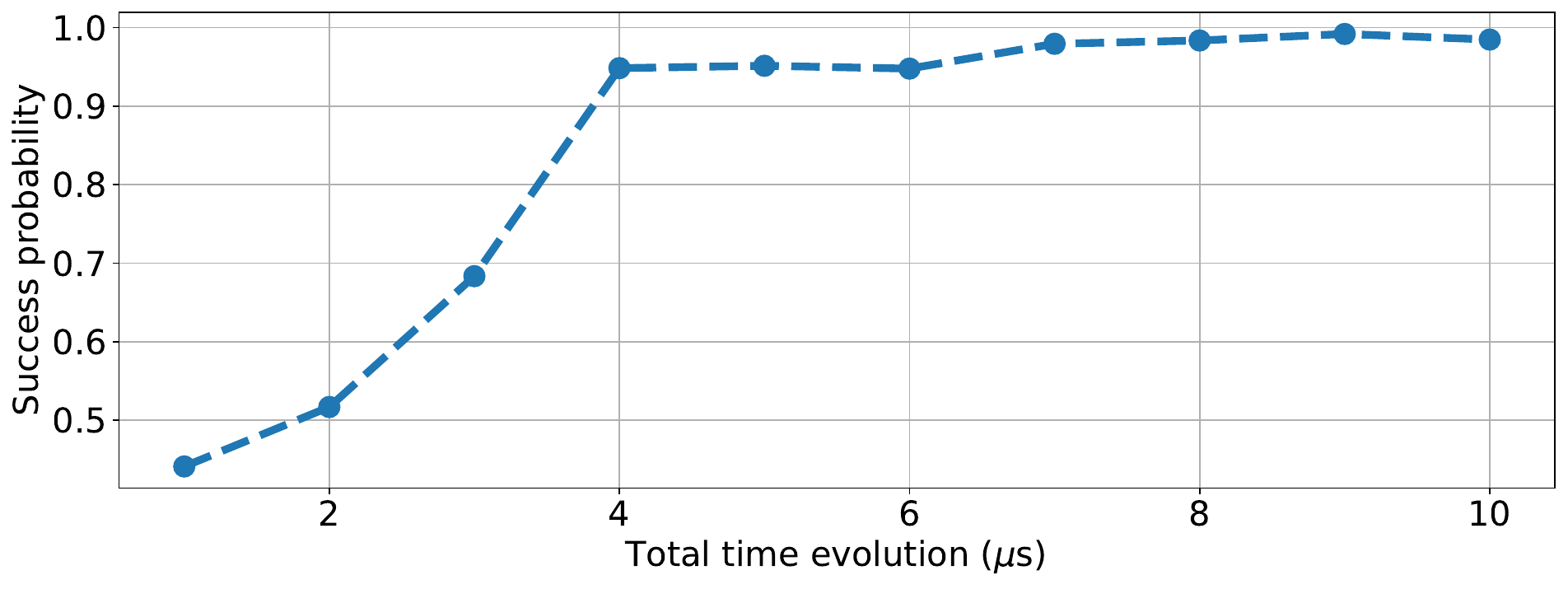}  
  \caption{ (Top) Graph encoded to solved the MIS problem. (Bottom) Success probability of solving the MIS problem with different time evolution using the pulse showed in figure \ref{fig:example_adiabatic}. All the figures were obtain using \textit{Pulser}.}
  \label{fig:qaa_results}
\end{figure}

In figure \ref{fig:qaa_results} we can see an example of time evolution on the adiabatic algorithm. Slower times gives better solutions.

\subsection{Variational Quantum Adiabatic Algorithm (VQAA) }

The approach used in this work is a Variational Quantum Adiabatic Algorithm (VQAA) based on \cite{pichler2018quantum}. The adiabatic algorithm is known to return the ground state for a sufficiently long time T. But, due to the limited decoherence times of current NISQ devices and analog quantum simulators, the time T for an optimal solution is not possible to implement. Finding a path with a time possible for implementation is of great relevance for the feasibility of the adiabatic approach. For example, the time evolution cannot be of 20 $\mu s$ because at the moment it is not possible to implement that time in real hardware.  Normally, you can do a sweep over the parameters, but this is computationally expensive. The idea of the VQAA is to find the correct parameters for the adiabatic evolution using a classical optimizer (in this case the ones from Scipy). 

We aim to find an optimized profile for the adiabatic evolution determined by the parameters $\Omega$, $\delta$ y $t$. The VQAA allows for a significant acceleration compared to the QAA with a linear adiabatic path \cite{schiffer2022adiabatic}, yet requires fewer parameters and measurements than the QAOA. As we can see in figure \ref{fig:vqaa_process}, different parameters give us different adiabatic paths, that may or not may be possible in real hardware. We aim to find the correct path that is possible to implement in real quantum hardware.

%%-----------------------------------------------------%%

\section{Hyperopt VQAA}

\subsection{Hyperopt}

To determine the parameters of the various pulses sent to the system, we employ an optimization technique called Hyperopt \cite{bergstra2013making}, an extension of Bayesian optimization. This approach efficiently searches for optimal parameters within a defined search space. Hyperopt replaces the optimization function from scipy used in the first version.

\subsection{Parameters}

Five parameters are defined: the rise time and fall time, representing the rise and fall times of the main pulse (Complex pulse used in the Scipy approach). Their total should not exceed the current coherence time of existing machines, set at 5 $\mu s$. Their intervals are between 16 (the minimum possible) and 2500 ns. $\Omega$, corresponding to the Rabi frequency, depends on the minimum and maximum distance between qubits in the register. The parameter bounds are dynamically calculated for each configuration. Finally, the initial and final detunings, $\delta_{0}$ and $\delta_{f}$, are the boundaries for the detuning values during the evolution of the second pulse. Their bounds are set between 0 and 8 rad/µs, with the maximum being 8.

\subsection{Relevance of results}

We introduce a measure to assess the relevance of results, which is a score used as the target value by Hyperopt during parameter optimization. The goal is to obtain high output numbers for the state(s) that represent MIS and low occurrences for all others. Additionally, we evaluate if a result is significant enough to be used, opting for the Gini coefficient. This metric assesses the disparity in the distribution of output configurations. While it does not guarantee a good or correctly distributed result, it serves as a good indicator when aiming for a few states to stand out from many others. It may not work effectively for a fully connected graph, where MIS will consist of a single node and be numerous.

Let \( G \) be the graph with \( N \) nodes, and \( C  \) the ensemble of \( M = 2^{N}\) possible binary configurations representing selected nodes (1 if selected, 0 otherwise).
The function \( f \)  evaluates the quality of a configuration \( C_i \) by summing all of its \( N \) bits noted \(C_{i,j}\) only if they form an independent set, noted as \( I(C_{i}) = 1 \) if the set is independent, 0 otherwise. The total is then divided by \( N \).

We then have:
\begin{equation}
\begin{gathered} f(C_i) = \frac{1}{N} \sum_{j=1}^{N} C_{i,j} \cdot I(C_i)\end{gathered}
\end{equation}

% \[
% f(C_i) = \frac{1}{N} \sum_{j=1}^{N} C_{i}(j) \cdot I(C_{i})
% \]

The Gini coefficient is defined as follows:
\begin{equation}
\begin{gathered} Gini(C) = 1 - \sum_{i=1}^{M}\left(\frac{\sum_{j=1}^{N}C_{i,j}}{N}\right)^2\end{gathered}
\end{equation}

Finally, the scoring function \( S \) aggregates the two previous functions, taking into account the quality of the found MIS and the quality of the distribution of results.

\begin{equation}
\begin{gathered} S(C) = \frac{\sum_{i=1}^{M}f(C_i)}{M} \times Gini(C)\end{gathered}
\end{equation}

A threshold is implemented: if the Gini coefficient is less than 1/3, the result is deemed irrelevant, and the score is nullified. This approach enables the scoring of a configuration without prior knowledge of the MIS. Initially, it maximizes the search for independent sets and subsequently increases their likelihood.

%%-----------------------------------------------------%%

\section{Machine Learning QAA (MLQAA)} 

\subsection{Use Case}

The primary bottleneck of the VQAA is the computation time within the parameter optimization loop. It is necessary to perform this operation for every new graph, significantly increasing the overall computational cost. To address this issue, we introduce graph machine learning. By accumulating sufficient data from previously executed VQAA runs, it becomes possible to associate a graph with its most probable parameters. This technique is akin to Meta Learning in classical machine learning.

With a properly trained model, the number of cycles executed by the QPU reduces from around ten to just one—the cycle needed to obtain the final result. These optimization cycles are replaced with a classical model inference, presenting a favorable tradeoff between total execution time and parameter precision.

Previous works conducted by Coelho et al. \cite{coelho2022efficient}, explored similar aspects to our study. 
While our approaches may exhibit similarities, it is important to note that our work was conducted independently, without prior knowledge of the results obtained by the preceding team. 
In comparison with their approach, we have chosen a different kind of model to obtain the graph embedding, and chose to do single target regression as opposed to their multi-target approach.

\subsection{Generation of the Data \label{sec:dataset}}
\subsubsection{Configurations}
 
The initial step involves generating graphs for training, aiming for sufficient diversity to enable the model to extrapolate to new structures. Additionally, an adequate number of graphs is essential to ensure meaningful training without inducing overfitting. Various structures, including lines, rectangles, triangles, triangular lattices, and hexagons, are generated.

The final parameter to manipulate is the spacing between qubits in the register for these configurations. All configurations are generated with spacings ranging from 6 to 11 $\mu m$. The benchmark for our solution is conducted on a set of 125 graphs generated in this manner.

\subsubsection{VQAA}

For each of these configurations, optimal parameters are generated using the VQAA function with Hyperopt. The optimization loops are numerous, averaging 50, to ensure the quality of the parameters. If the score of the best parameter set is not null due to a too-low Gini coefficient, it is then stored along with the parameter set. For low scores, a second pass of VQAA with increased depth is applied, and the new parameters are saved if the score is improved.

 \subsubsection{Pre-processing Data}
 
The model is trained on the graph formed by atoms in the register. The adjacency matrix between all qubits is computed as the inverse of their pairwise squared distance, representing the interaction strength between these atoms. The exponent's value is experimentally chosen through parameter optimization. Each of these adjacency matrices is associated with the five parameters calculated by VQAA, as discussed in the preceding section.

We employ the PyTorch Geometric framework for data and model preparation. The model's input is the adjacency matrix split into two parts to adhere to framework conventions. First, an edge indices array, denoting the various pairwise node connections, is of dimension \((2, N \times (N-1))\) for \(N\) the number of nodes, with edges appearing in both directions to represent a non-directional graph. Second, an edge weights array, corresponding to values in the adjacency matrix, is of size \((1, N \times (N-1))\). Finally, the framework requires node weights, which are absent in this case, thus taking a value of 1 for the entire \( (1, N)\) dimensional array.

The final step involves aggregating all graphs into a data loader, facilitating the provision of batches of predefined size. This technique optimizes model training performance while allowing for passive regularization.

\subsection{Model}
\subsubsection{Model Choice}

The chosen model implements graph convolution layers operating on the principle of message passing. Specifically, the model comprises 5 Graph Convolutional Network layers (GCN) \cite{GCN} with a hidden dimension of 128, each associated with a dropout of 0.1 - for training purposes only - and a ReLU activation function. These convolutional layers are then connected to a graph-level add pool, yielding an embedding for each graph. This embedding is subsequently utilized in the final layer, a linear layer with a single output.

\begin{figure}[h]
\centering
     \includegraphics[width=0.95\linewidth]{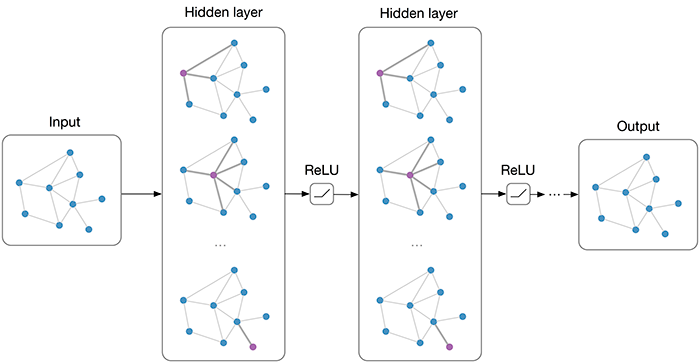} 
  \caption{Multi-layer Graph Convolutional Network (GCN) performing multiple graph convolutions to obtain node embeddings. Image from \cite{GCN}} 
  \label{fig:gcn}
\end{figure}

\subsubsection{Training}

A separate network is trained for each target value. This decision is based on the likelihood that each parameter relies on different features for prediction. Employing a multi-target regression with all five parameters as outputs would necessitate a larger and more complex model, yielding uncertain results. It seems more reasonable to conduct training for each parameter individually, followed by aggregating the 5 trained models into a single entity.

Each model undergoes training with batches of 64 elements. The Adam optimizer is employed, with a learning rate ranging between \( 0.01\) and  \(0.0001\), depending on the target parameter.

%%-----------------------------------------------------%%

\section{Results}

\subsection{QAA limitations}

\begin{figure}[h]
\centering
      \includegraphics[width=1\linewidth]{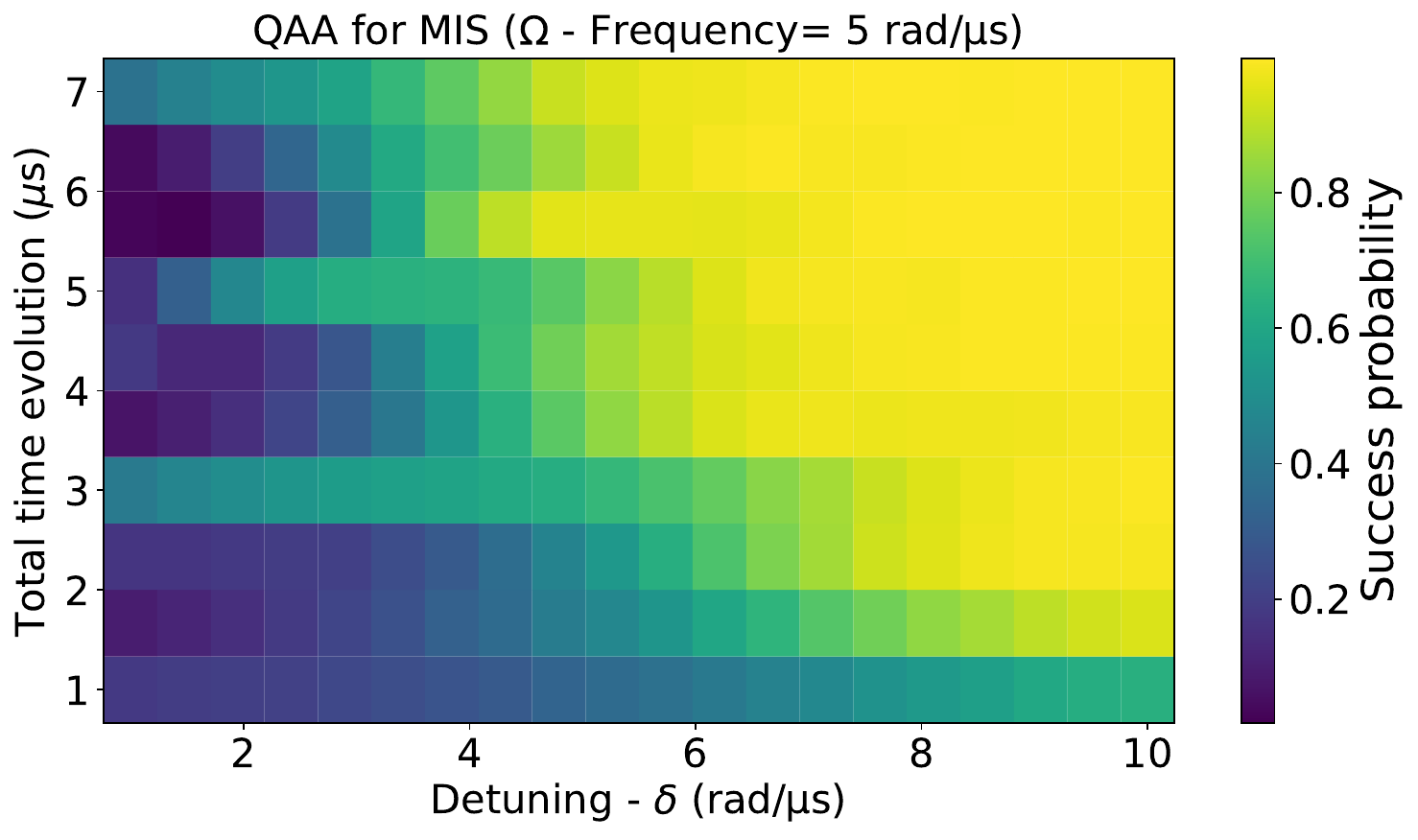}    
  \caption{Results for the graph shown in figure \ref{fig:qaa_results_encoding}. Sweeping different detuning values ($\delta$) with a constant frequency ($\Omega$) and different time evolution for the adiabatic pulse. The different colors determine the probability of success of QAA.}
  \label{fig:qaa_result_2}
\end{figure}

Working with QAA quickly becomes difficult, given the search for the right parameters to run the adiabatic evolution. The frequency($\Omega$) and detuning($\delta$) would change drastically from one problem to the other as we can see in figure \ref{fig:qaa_results} and figure \ref{fig:qaa_result_2}. There, we show a grid search approach which is especially slow but very accurate for the solution of the graph shown in figure \ref{fig:qaa_results}. Further more the search for the parameters is relatively long and tedious, those difficulties would find themselves only enhanced with the scaling of the problem. The search space would become much larger and with the time constraint of the physical device we would require multiples runs, which means we would need to reset the atoms which would further improve the time of the parameters search.

To alleviate the limitations of the QAA solution we used a hybrid approach where the parameters are found trough a classical optimization process.

\begin{figure}[h]
\centering
    \includegraphics[width=1\linewidth]{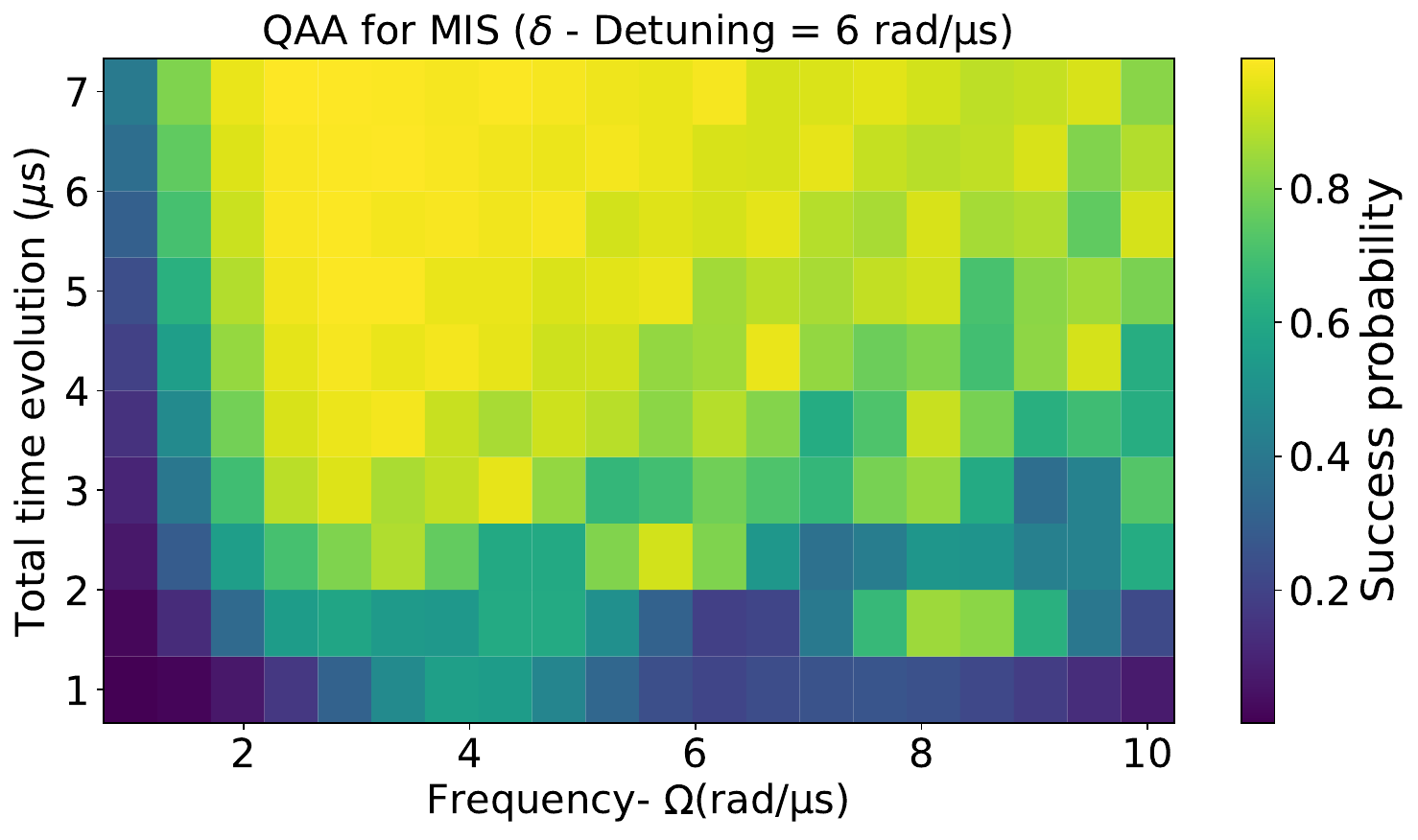}    
  \caption{Results for the graph shown in figure \ref{fig:qaa_results_encoding}. Sweeping different frequencies ($\Omega$) with a constant detuning ($\delta$) and different time evolution for the adiabatic pulse. The different colors determine the probability of success of QAA.}
  \label{fig:qaa_result_1}
\end{figure}

\subsection{Scipy VQAA}

The VQAA was performed with two different type of adiabatic sequences defined by: 

\begin{itemize}
    \item Simple Sequence: \texttt{InterpolatedWaveform}
    \item Compelex Sequence: \texttt{RampWaveform}
\end{itemize}

\begin{figure}[h]
\centering
      \includegraphics[width=0.6\linewidth]{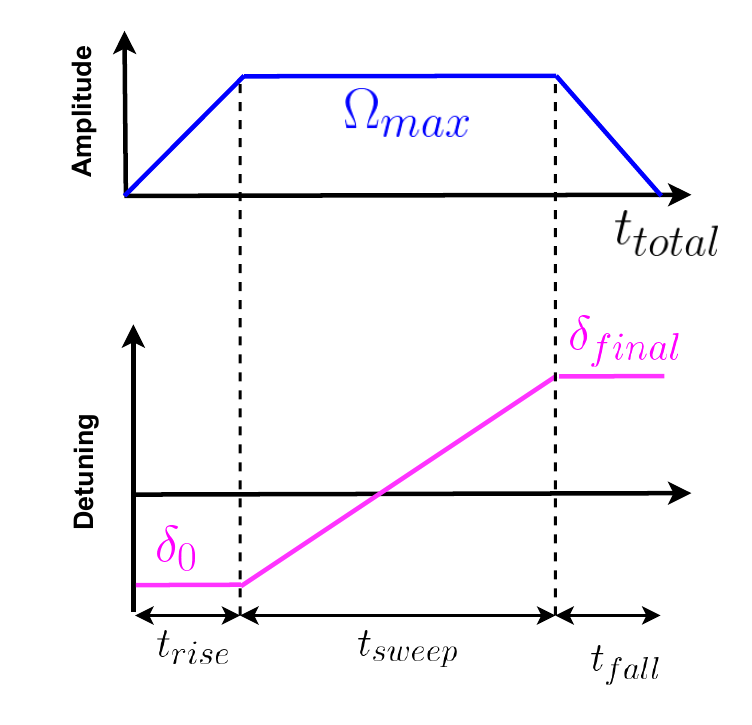}  
  \caption{Complex pulse using \texttt{RampWaveform}. }
  \label{fig:complex_sequence}
\end{figure}

An example of a simple sequence is shown in figure \ref{fig:example_adiabatic} and of the complex sequence in figure \ref{fig:complex_sequence}. When the simple sequence is used the parameters to optimized are the following:  Frequency ($\Omega$), Detuning ($\delta$) and Time ($t$).

In the case of using the complex sequence the number of parameters to optimized increases: Frequency ($\Omega$), Initial Detuning ($\delta_t$), Final Detuning ($\delta_f$), Time rise ($t_{rise}$) and Time fall ($t_{fall}$).

%\begin{figure}[h]
%      \includegraphics[width=0.8\linewidth]{figure/VQAA_diagram.png}  
 % \caption{Workflow for the VQAA.The MIS problem is encoded into the atomic array with an Unit-Disk approach. The whole quantum system evolves under a programmable laser drive that is characterized by the frequency ($\Omega(t)$) and detuning ($\delta(t)$). A measurement is performed which give the independent set of the graph. Then a classical optimizer uses the results to update the parameters ($\Omega, \delta, t$). Image from \cite{pichler2018quantum}. }
 % \label{fig:vqaa_diagram}
%\end{figure}

\begin{figure}[h]
\centering
      \includegraphics[width=0.4\linewidth]{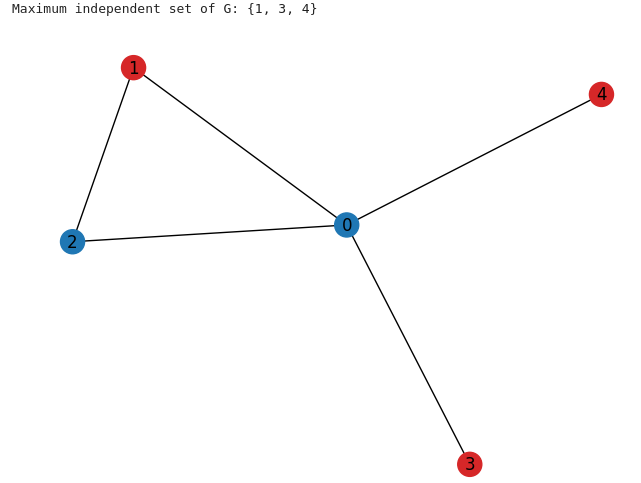}  
      \includegraphics[width=0.4\linewidth]{figure/graph_qaa.png}  
  \caption{(Left) Encoding into atomic array. (Right) Graph to solve the MIS problem.}
  \label{fig:qaa_results_encoding}
\end{figure}

For the minimizers, \texttt{scipy} was used with the standard optimizer   \texttt{Nelder-Mead} or  \texttt{COBYLA}. In figure \ref{fig:qaa_results_encoding} we can see the results of VQAA for the graph of 5 nodes shown in figure \ref{fig:qaa_results_encoding}. 

\begin{figure}[h]
\centering
     \includegraphics[width=0.8\linewidth]{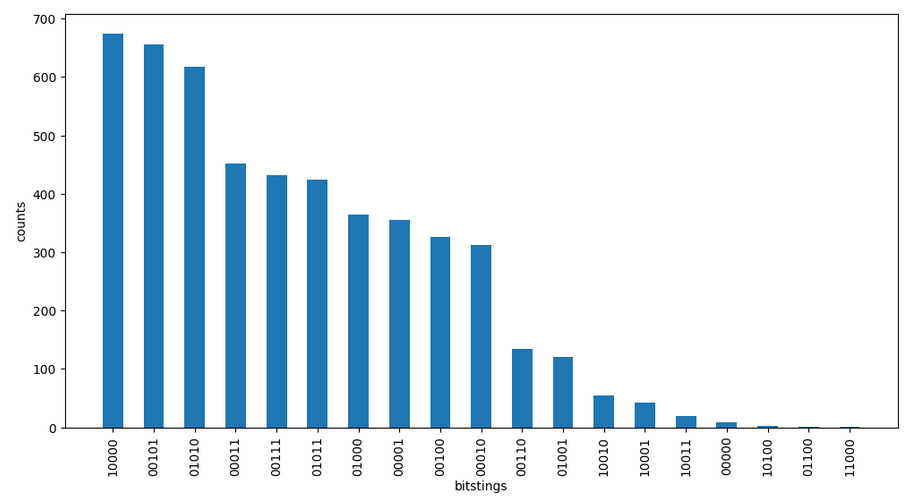} 
  \caption{Results using the VQAA for the problem shown in figure \ref{fig:qaa_results_encoding} }
  
  \label{fig:results_vqaa_5_atoms}
\end{figure}

Simple sequences was used with the \texttt{Nelder-Mead} optimizer with 4 repetitions of the optimization choosing the best one. The optimal time evolution found by the VQAA is 5429 ns with $\Omega = 1.86$ and the $\gamma= 3.02$. The histogram of the different solutions is shwon in figure \ref{fig:results_vqaa_5_atoms}. 

\begin{figure}[h!]
\centering
      \includegraphics[width=0.4\linewidth]{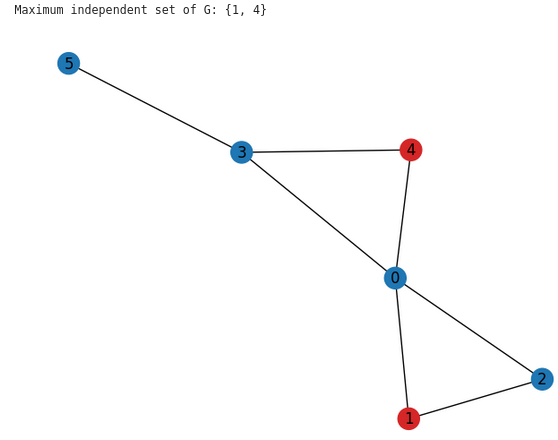}  
     \includegraphics[width=0.4\linewidth]{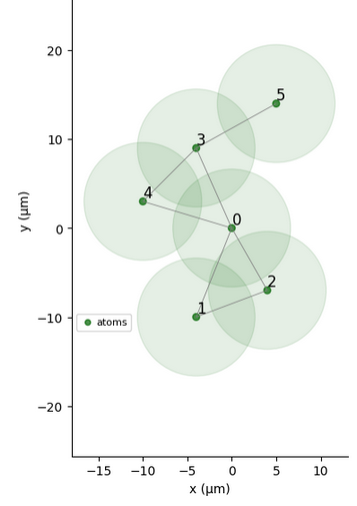} 
  \caption{Encoding of a 6 nodes graph into a atomic array. The MIS of this graph was found with VQAA, using a complex sequence. }
  \label{fig:encoding_6_atoms}
\end{figure}

\begin{figure}[h!]
\centering 
     \includegraphics[width=0.8\linewidth]{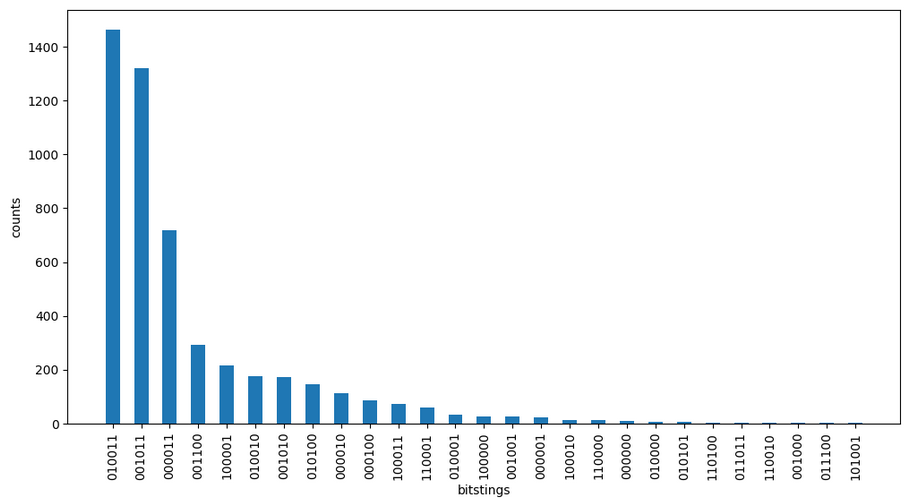} 
  \caption{Results using the VQAA for the problem shown in figure \ref{fig:encoding_6_atoms} }
  
  \label{fig:results_vqaa_6_atoms}
\end{figure}

The mapping of a 6 nodes graph to an atomic array can be seen in figure \ref{fig:encoding_6_atoms}. The results of applying VQAA to this graph are shown in in figure \ref{fig:results_vqaa_6_atoms} . Complex sequences were used with the \texttt{Nelder-Mead} optimizer with 5 repetitions of the optimization choosing the best one. The optimal time evolution found by VQAA is 6258 ns for $t_{rise}$ and 3011 for ($t_{fall}$)  with $\Omega = 7.56$ and the $\delta_{initial} = 3.43$ and $\delta_{final} = 3.46$.  The histogram of the different solutions is shown in figure \ref{fig:results_vqaa_6_atoms}.

\begin{figure}[h]
\centering
      \includegraphics[width=0.6\linewidth]{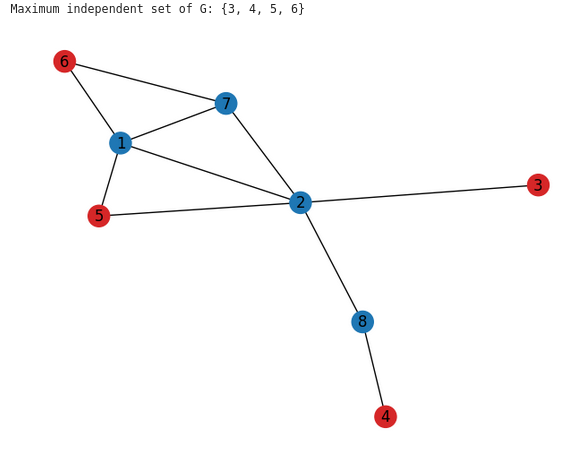}  
     \includegraphics[width=0.6\linewidth]{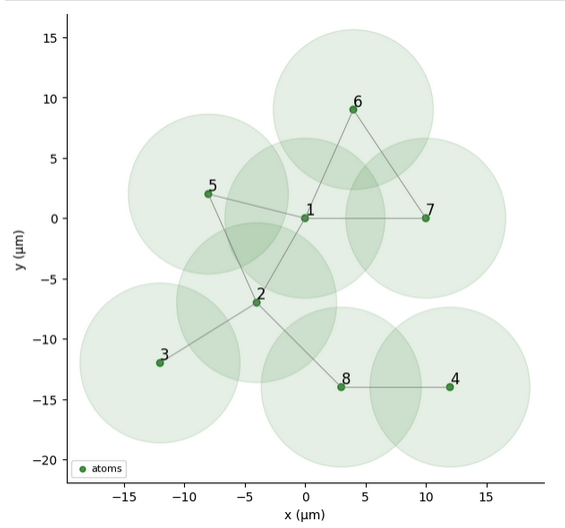} 
  \caption{Encoding of a 8 nodes graph into a atomic array. The MIS of this graph was found with VQAA. }
  \label{fig:encoding_8_atoms}
\end{figure}

An eight nodes graph was encoded into an atomic array as shown in figure \ref{fig:encoding_8_atoms}. A simple sequence was used  with the \texttt{Nelder-Mead} optimizer with 5 repetitions of the optimization choosing the best one. The optimal time evolution found by VQAA is 4110 ns with $\Omega = 4.7$ and the $\gamma = 4.69$.

%\begin{figure}[h]
%\centering
%\includegraphics[width=1\linewidth]{figure/rho_Vs_prob_more_data.pdf} 
%  \caption{
%Success of probability using VQAA as a function of total number of MIS solutions (MIS degeneracy, $D_{MIS}$) normalized by the graph size. \colorbox{red}{I need to explain this plot} }
 % \label{fig:rho_Vs_probs}
%\end{figure}

In all the cases study, the optimal solution was found in the list of best solutions provided by the VQAA. Also, the time evolution for the adiabatic process in all instances are in range with the future improvements of PASCAL QPU, which currently is on the order of 4 $\mu s$. Our solutions are on the order of 4 to 9  $\mu s$ which are possible to improve with modifications in the parameters of the optimizer. Even though we couldn't explore the algorithm's behavior in bigger instances given the computational limitations, we have shown the feasibility of using VQAA for quantum solve of the MIS problem which is the solution for the Max Clique given the complementary graph.

At the moment of increasing the size of the graph the runtime increases significantly and the \texttt{scipy} optimizer becomes unstable. We introduced a more sophisticated optimizer using Hyperopt, improving the results and sizes of graphs possible to solved. 

%Further improvements are necessary in the classical optimizer (explore different optimizer and options of the minimizes) and pulse sequences (used more complex pulses). 

\subsection{Hyperopt VQAA}

VQAA is applied to the dataset comprising 125 graphs, as described in the dataset section (\ref{sec:dataset}) with varying numbers of optimization rounds to observe differences. Scores are collected and normalized by the size of the MIS. This normalization yields a value between 0 and 1, representing the probability of obtaining an optimal solution to the problem. Different number of rounds, from 10 to 500, are ran to compare their results.

\begin{figure}[h!]
\centering
\includegraphics[width=0.95\linewidth]{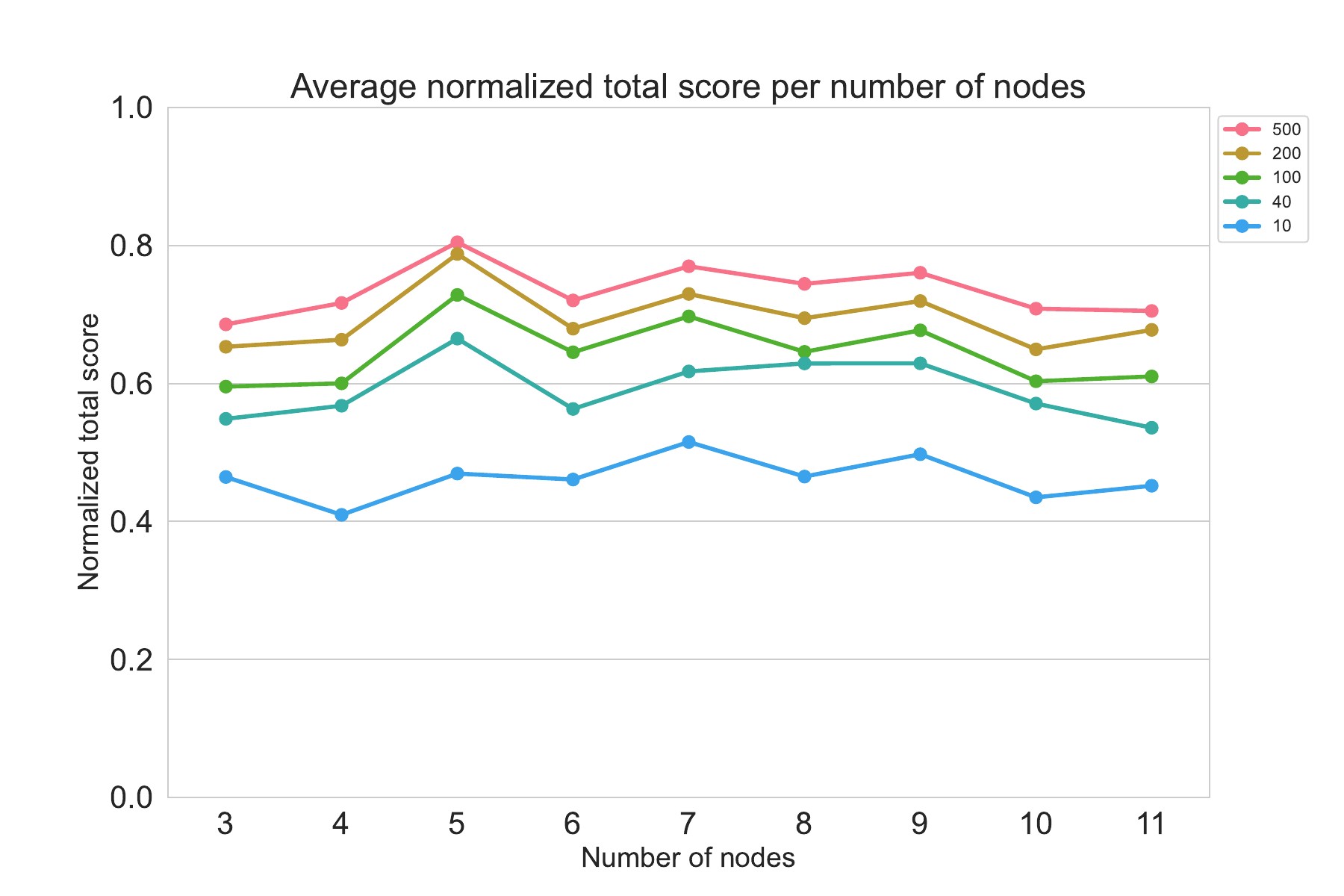}  
    \caption{VQAA results per graph size. The different lines are the rounds of optimiszation of each VQAA.}
    \label{fig:vqaa_results_nodes}
\end{figure}

\begin{figure}[h]
\centering
\includegraphics[width=0.95\linewidth]{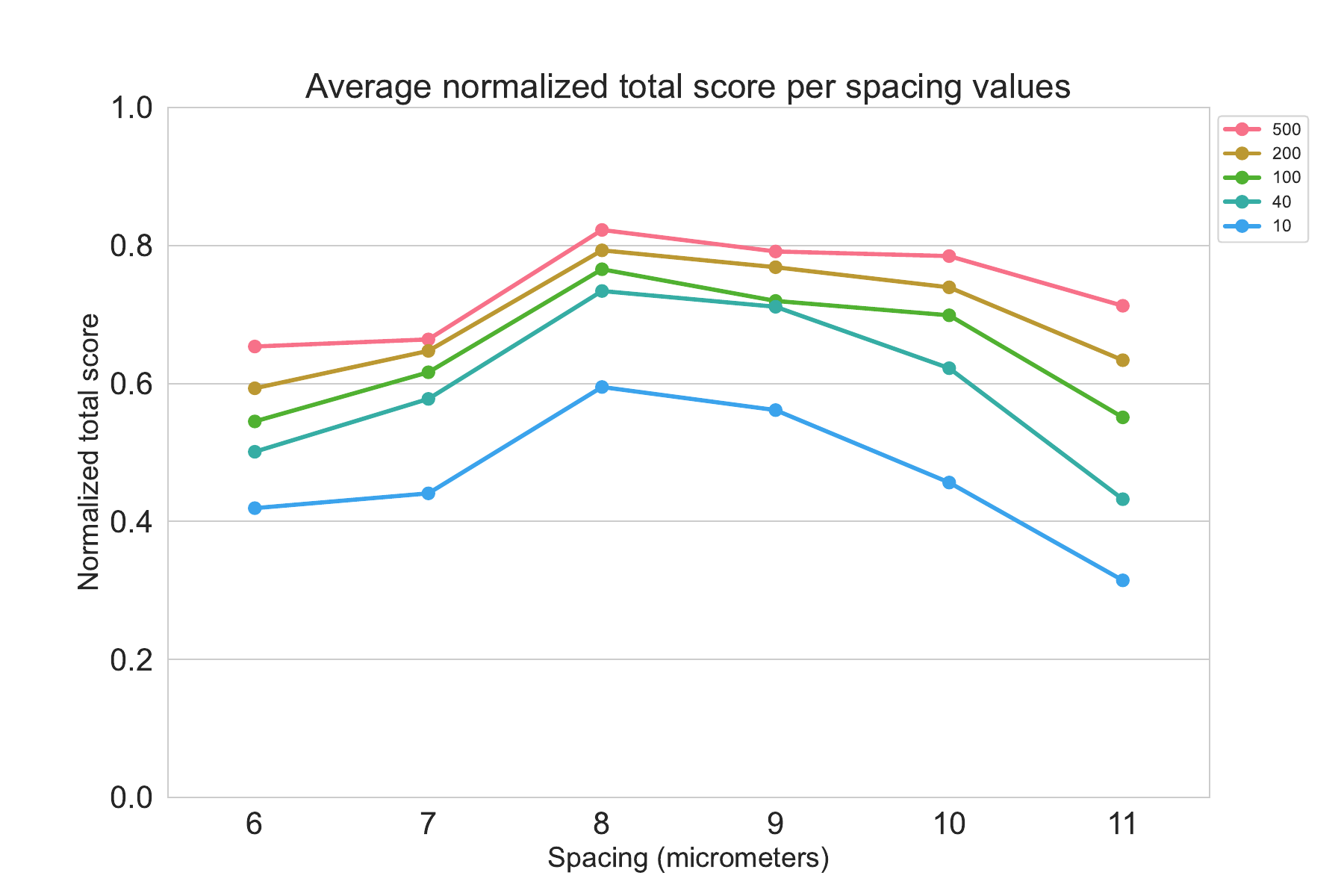}  
    \caption{VQAA results per spacing values between the register's atoms.}
    \label{fig:vqaa_results_spacing}
\end{figure}

The scores are averaged by number of nodes in the graph, and by spacing between the atoms to give insights on the performance of VQAA in those different scenarios. In figure \ref{fig:vqaa_results_nodes}, it's observed that the normalized score logarithmically increases with the number of optimization rounds. Also, as it is shown in figure \ref{fig:vqaa_results_spacing}, the spacing between qubits is also crucial, especially with a low number of optimization rounds, as the parameter convergence window seems more restricted. Notably, for a spacing of 8 mm, an average score demonstrates a certain efficiency of the process, even for a relatively large graph and with only 10 iterations.

\subsection{MLQAA}

\begin{figure}[h!]
\centering
\includegraphics[width=0.95\linewidth]{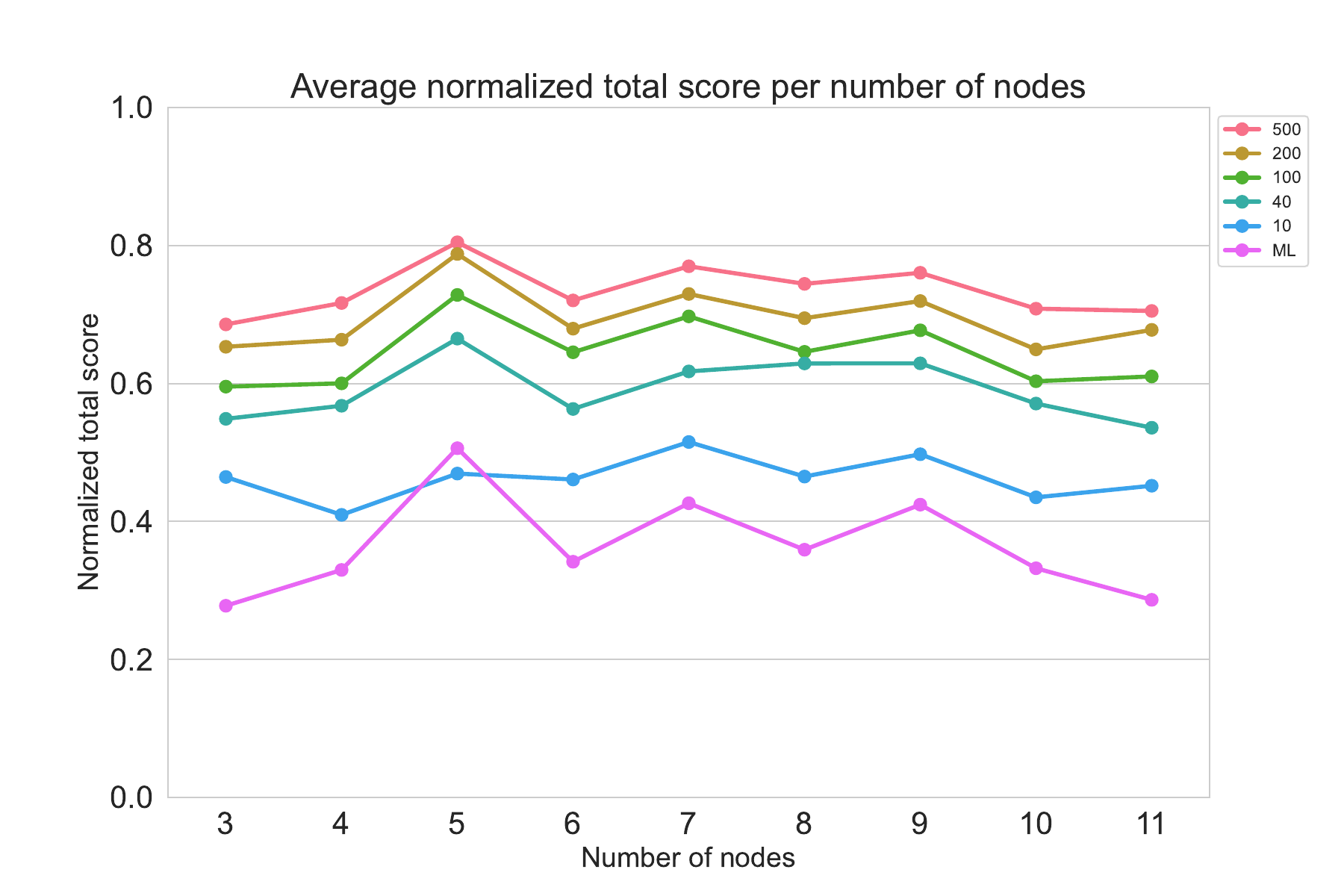}  
    \caption{MLQAA results per graph size.}
    \label{fig:ml_nodes}
\end{figure}

\begin{figure}[h!]
\centering
\includegraphics[width=0.95\linewidth]{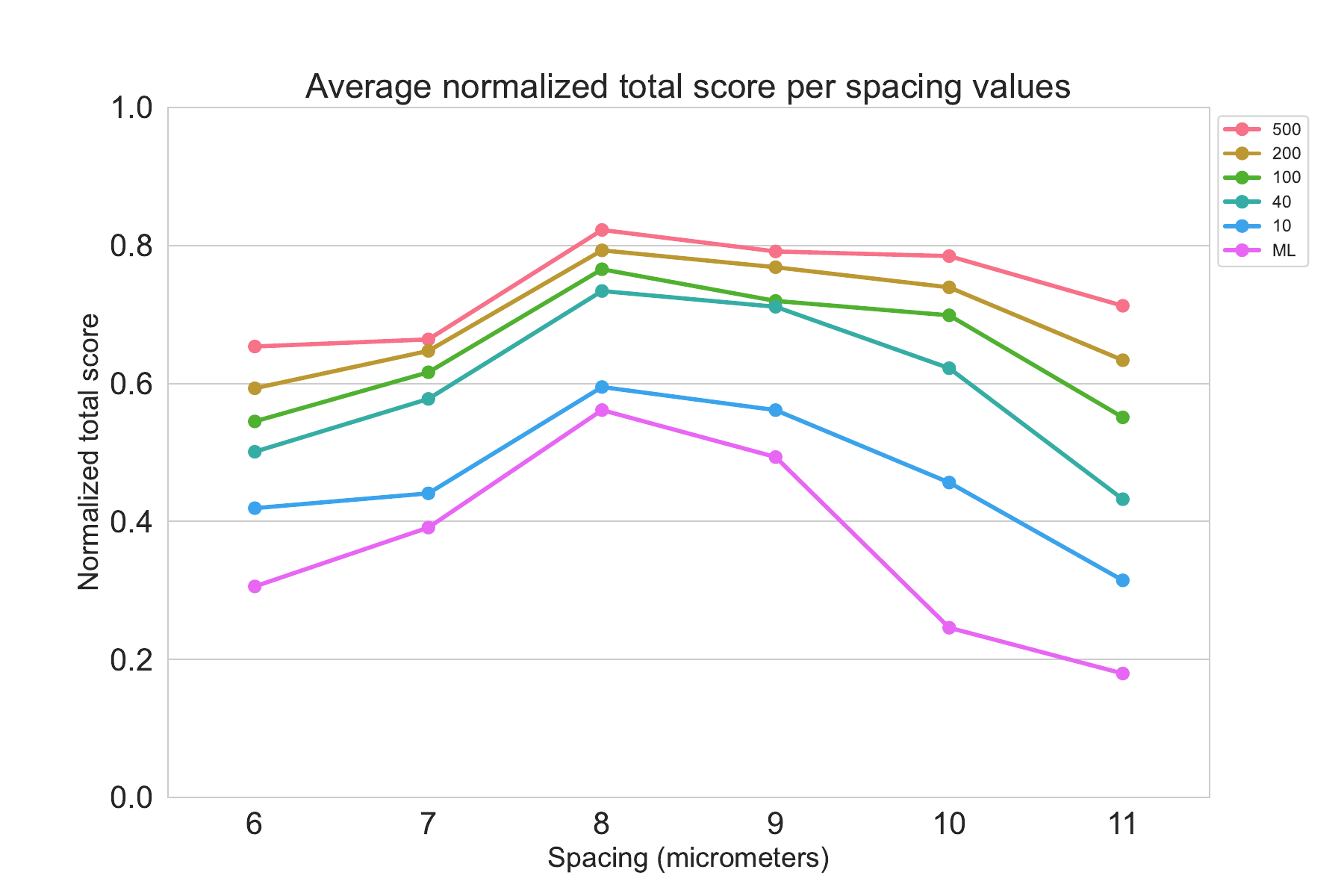}  
    \caption{MLQAA results per spacing values between the register's atoms.}
    \label{fig:ml_spacing}
\end{figure}

The final Mean Average Percentage Error (MAPE) values obtained for each parameter model are shown in table \ref{table:parameters_results}.

\begin{table}[h!]
\centering
\begin{tabular}{ |c|c|c| } 
 \hline
 Parameter & MAPE   \\ 
 \hline
 Rise time ($t_{0}$)  & 22.3 \\ 
\hline
 Fall time ($t_{final}$) & 18.3 \\ 
 \hline
 Frequency ($\Omega$) & 21.4 \\ 
 \hline
 Initial detuning ($\delta_{0}$) & 16.7  \\ 
  \hline
 Final detuning ($\delta_{f}$) & 21.5  \\ 
 \hline
\end{tabular}
  \caption{MAPE for the different parameters using MLQAA}
\label{table:parameters_results}
\end{table}

Objectively, these values are not particularly good, especially considering the impact of a small error on performance, particularly in cases with significant spacing between qubits. However, this outcome is not surprising. The results obtained by VQAA, even with 500 iterations, still exhibit considerable variability. In a given graph, where the operation was repeated 50 times, the value of the $\Omega$ parameter ranged from 10 to 15. Moreover, the number and variety of graphs are insufficient to ensure quality training.

Despite these challenges, the results are still encouraging. As shown in figures \ref{fig:ml_nodes}, \ref{fig:ml_spacing}, the model manages to identify interesting features in the proposed embedding method, as the MLQAA results are only slightly below those of VQAA at 10 rounds. In some use cases like molecular docking where multiple near-optimal results are expected, the trade-off between computation time and performance can already be beneficial.

With a larger and more diverse dataset, along with an increased number of iterations to generate target parameters, this type of model, based on these results, could become a crucial tool in addressing VQAA bottlenecks.  It would be interesting to combine this approach with the Multi-Target approach proposed by \cite{coelho2022efficient}. Finally, establishing a hybrid system that leverages machine learning to provide parameters used as initial values for VQAA could reduce the number of required searches.

\subsection{Molecular Docking}

%\colorbox{red}{Section needs to improve}
%\colorbox{red}{we need to shown a clear working example}

The MLQAA method presents particularly interesting results for molecular docking. A normalized score of 0.8 is considered excellent, indicating that the most crucial configurations to evaluate with the scoring function (\ref{sec:scoring_function}) are highly likely to be among the top 10 configurations.

\begin{figure}[h!]
\centering
  \includegraphics[width=0.48\linewidth]{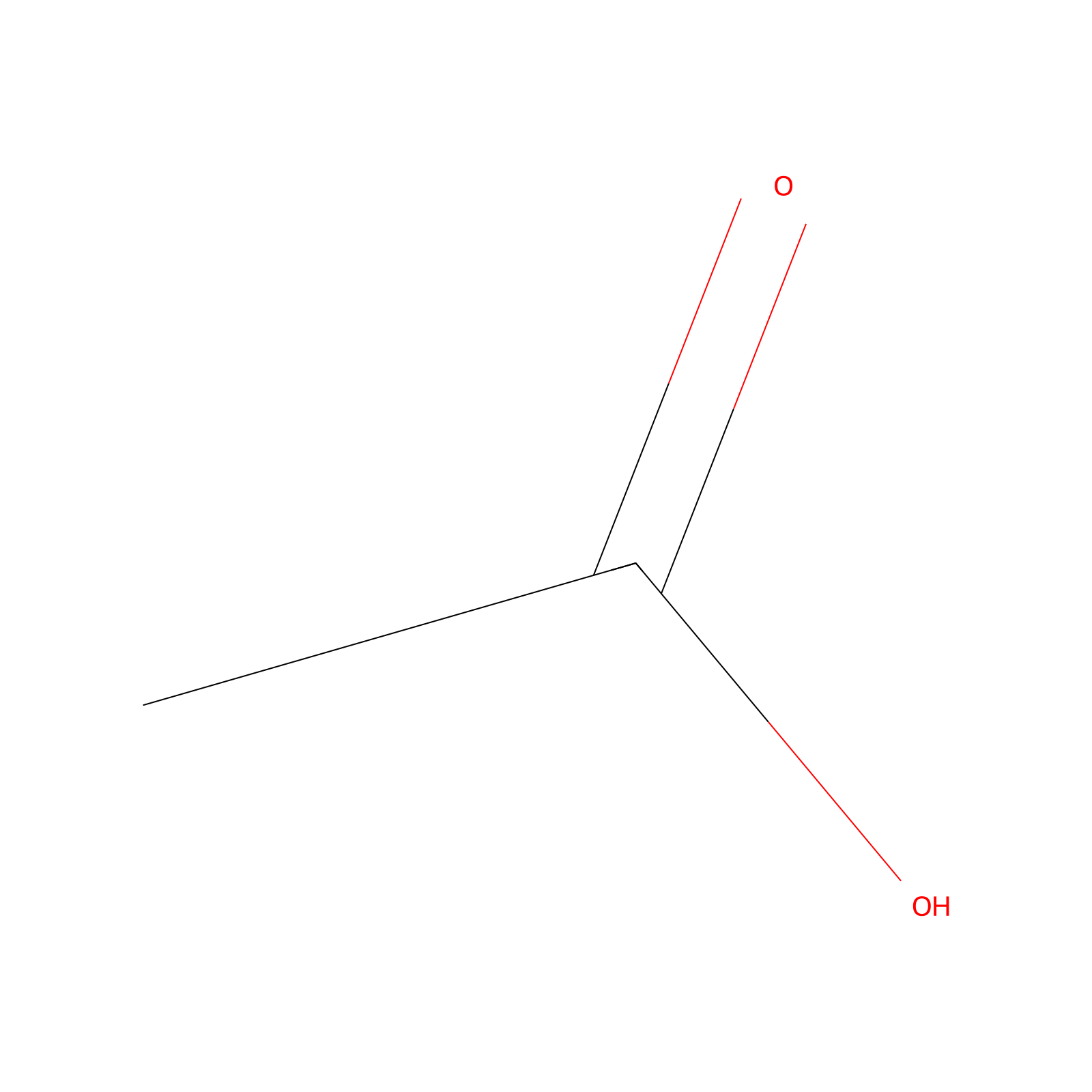}
  \includegraphics[width=0.48\linewidth]{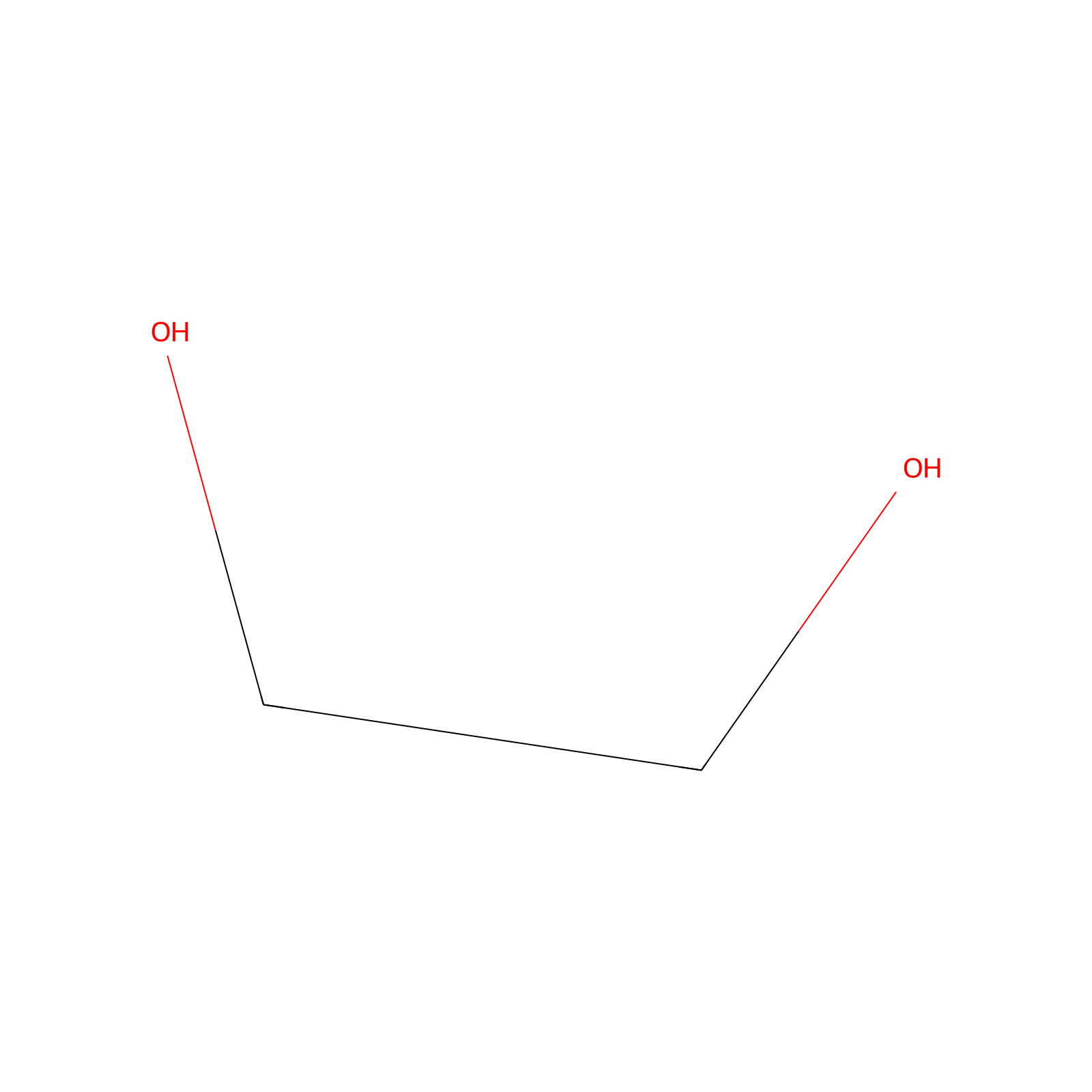}
\caption{(Left) Acetic acid (Right) Ethylene glycol}
\label{fig:example_molecules}
\end{figure}

\begin{figure}[h!]
\centering
  \includegraphics[width=0.48\linewidth]{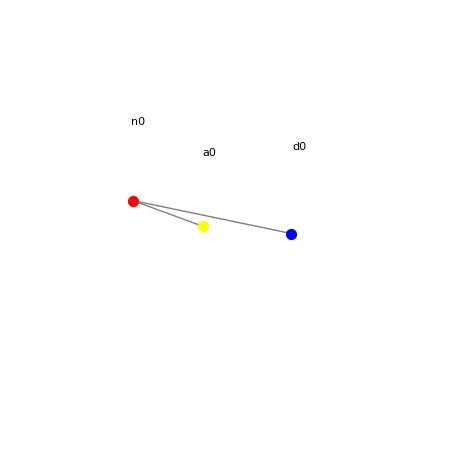}
  \includegraphics[width=0.48\linewidth]{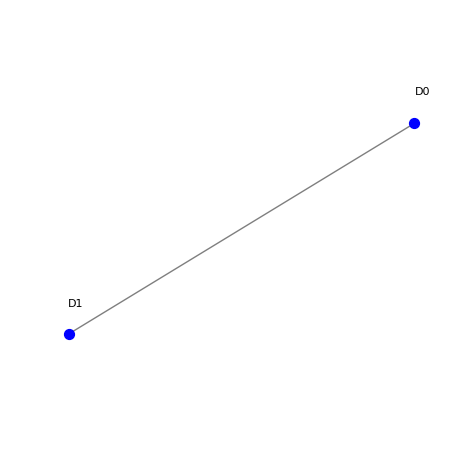}
\caption{(Left) Acetic acid pharmacophore points (Right) Ethylene glycol pharmacophore points}
\label{fig:example_pharmacophore_points}
\end{figure}

\begin{figure}[h!]
\centering
  \includegraphics[width=0.8\linewidth]{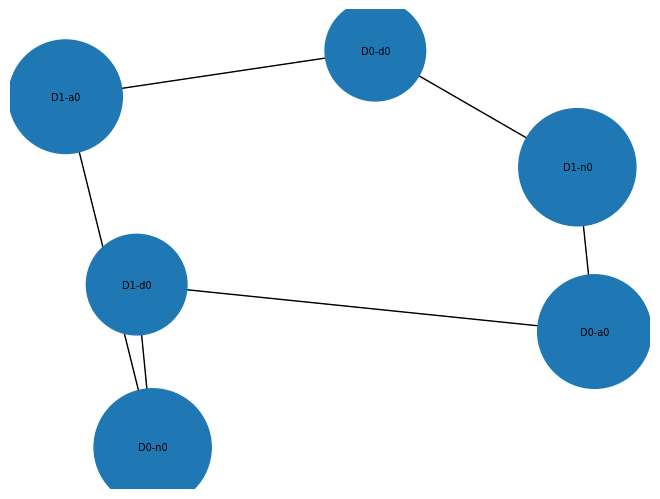}
\caption{Binding interaction graph of Acetic acid and Ethylene glycol.}
\label{fig:example_binding_graph}
\end{figure}

A proof of concept is conducted with two small molecules capable of interacting with a molecular docking site, namely acetic acid and ethylene glycol (figure \ref{fig:example_molecules}). Their pharmacophore points are extracted, and the binding interaction graph is generated (figures \ref{fig:example_pharmacophore_points}, \ref{fig:example_binding_graph}). 

\begin{figure}[h!]
\centering
  \includegraphics[width=0.5\linewidth]{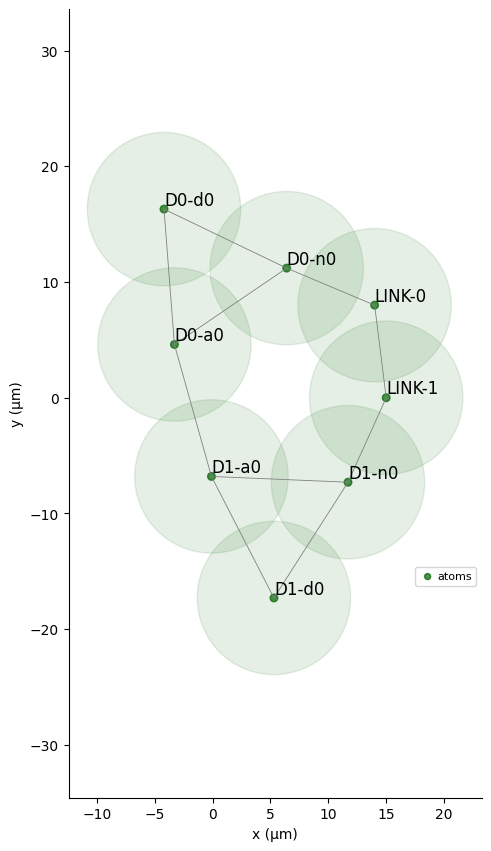}
\caption{Conjugate graph mapped to register, with added quantum links.}
\label{fig:example_register}
\end{figure}

\begin{figure}[h!]
\centering
  \includegraphics[width=0.9\linewidth]{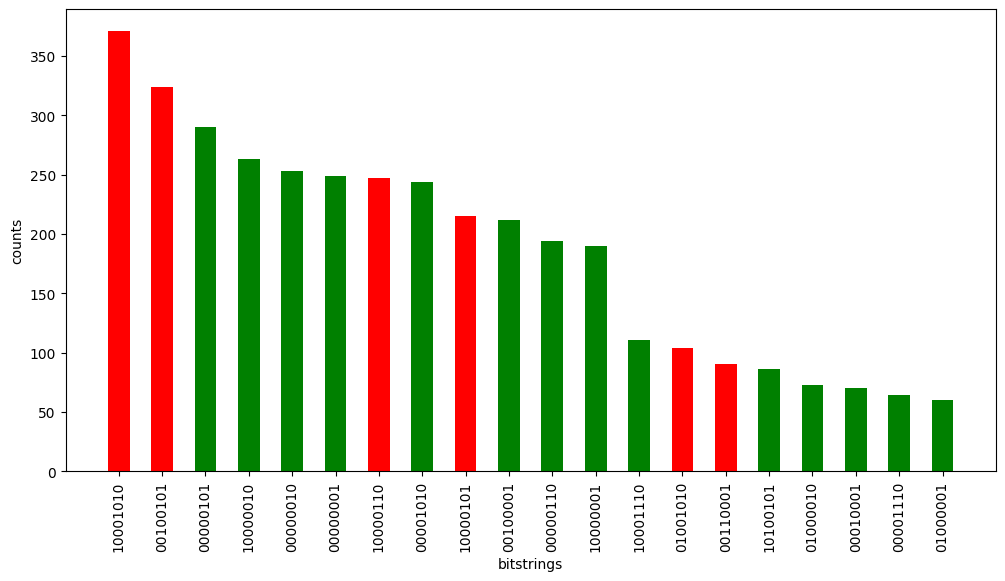}
\caption{10 rounds VQAA results for the Molecular Docking. Actual MIS are in red.}
\label{fig:example_vqaa}
\end{figure}

\begin{figure}[h!]
\centering
  \includegraphics[width=0.9\linewidth]{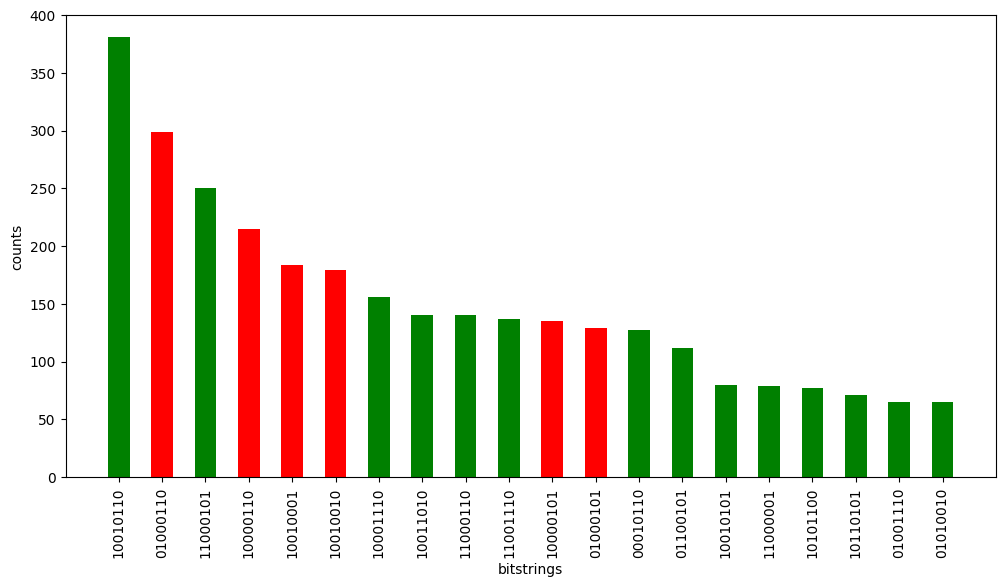}
\caption{MLQAA results for the Molecular Docking. Actual MIS are in red.}
\label{fig:example_mlqaa}
\end{figure}

\begin{figure}[h!]
\centering
  \includegraphics[width=0.9\linewidth]{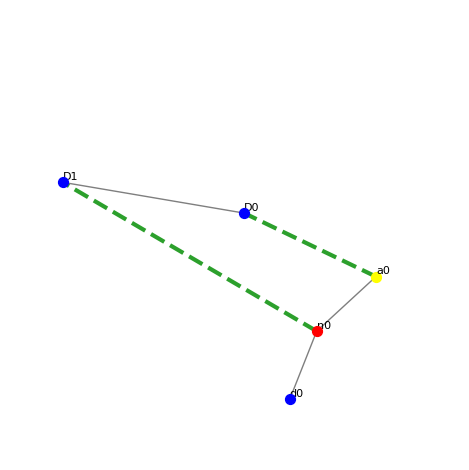}
\caption{Resulting docking between the two molecules pharmacophore points, corresponding to the found Max Clique in the binding interaction ghraph.}
\label{fig:example_binding}
\end{figure}

Quantum links (\ref{sec:links}) need to be added to the conjugate of the graph so that it can be mapped into an atom register (figure \ref{fig:example_register}). The VQAA algorithm with 10 rounds and MLQAA are applied. We can observe the histograms in figures \ref{fig:example_vqaa} and \ref{fig:example_mlqaa}. The obtained results, in a real-world scenario, allow for the selection of numerous promising candidates for further testing with a scoring function. For this proof of concept, we only select the highest value, resulting in the docking shown in figure \ref{fig:example_binding}.

The graphs used here may not be representative of real interaction graphs between molecules, which are at least twice as large for the smallest ones. Nevertheless, the algorithm will operate similarly on larger graphs, provided that quantum hardware evolves to extend coherence times, enabling better convergence on larger graphs. Additionally, in many cases there is prior knowledge of the docking area of the receptor molecule, which greatly decreases the size of the resulting binding interaction graph.

%%-----------------------------------------------------%%

%\section{Benchmark}

%\subsection{QAOA}

%\subsection{Tensor Network}

%%-----------------------------------------------------%%
\section{Discussion}

Previous works with molecular docking and quantum computing is presented in \cite{banchi2020molecular}. In this work, we have taken the same process to express the docking configurations as a graph problem. The main difference is that the solver is based on Gaussian Boson Sampling (GBS), which uses photonic quantum devices, to decrease the search space for max cliques of the system and later use greedy shrinking and expansion with local search to find the correct configurations.

The hybrid combination of quantum and classical process is similar to our approach where the classical part is the optimization or training of the ML model. However, our approach is more direct in solving the problem given the natural way to solve the MIS problem (in which the max clique is the complementary solution) with neutral atoms devices. As a result, our proposed methodologies make molecular docking a natural problem to solve with these devices.

Given our computational limitations, we were not able to compare a full molecular docking problem in a real-world application with our methods. But, the proof of concept presented showcases the feasibility and potential of our methodologies. Additionally the study with VQAA and MLQAA are consistent with similar works \cite{coelho2022efficient,ebadi2022quantum}.

\section{Conclusions}

% We have simplified the problem by identifying a probable docking site, which has drastically reduced the size of the graph. This method is also used in classical simulations. 
% These results serve as a proof of concept that neutral atom devices can predict molecular docking configurations.
In this work, the developed proof of concept demonstrates the feasibility of solving the molecular docking problem using neutral atoms. We have mapped the docking problem to the Max clique, which is complementary to the Maximum Independent Set problem. The use of neutral atoms provides a natural solution for this problem. This marks a new practical application for quantum devices based on neutral atoms. However, the current implementation bottleneck lies in embedding the problem into the register, presenting several challenges. Given the technology's current stage, we are confident this bottleneck will be overcome in the coming years.

We conducted a study on general graphs using various methods, Scipy VQAA, Hyperopt VQAA, and MLQAA. The main idea behind all these approaches is to find the right parameters for the adiabatic evolution of the system. Scipy, as a classical optimizer, produced accurate results for small graphs. However, when the size of the graph was increased, the optimization process took too much time, and the results became unstable. On the other hand, we observed an improvement in the results when using Hyperopt as the classical optimizer, even for different sizes and configurations of graphs. MLQAA proved to be the most promising implementation, delivering consistent and faster results. 

Our results highlight the potential use of NISQ devices based on neutral atoms for the drug discovery process. 

%%-----------------------------------------------------%%

\section{Acknowledgments}

We would like to acknowledge that the initial development of this work was created during the PASQAL hackathon, The Blaise Pascal [Re]generative Quantum Challenge. We are grateful for our multiple mentor's involvement, advice, and eagerness to help us learn in this hackathon. Additionally, we extend our thanks to Lorenzo Cardarelli and Elie Bermot for their insightful discussions and suggestions.

\bibliographystyle{plainnat}
\bibliography{ref}

\end{document}